\documentclass[useAMS,a4,usenatbib]{mn2e}
\input psfig.sty

\usepackage{epsfig}
\usepackage{psfig}

\def \chisq {\ifmmode \chi^2 \else $\chi^2$ \fi} 
\def \spose#1{\hbox to 0pt{#1\hss}} 
\def \lta{\mathrel{\spose{\lower 3pt\hbox{$\sim$}}\raise 2.0pt\hbox{$<$}}}
\def \gta{\mathrel{\spose{\lower 3pt\hbox{$\sim$}}\raise 2.0pt\hbox{$>$}}}

\def \kms {\ifmmode \,\rm km\,s^{-1} \else $\,\rm km\,s^{-1} $ \fi }
\def \kpc {\ifmmode {\rm~kpc} \else ${\rm~kpc}$\fi} 
\def \pc {\ifmmode {\rm~pc} \else ${\rm~pc}$ \fi } 
\def \Gyr {\ifmmode {\rm~Gyr} \else ${\rm~Gyr}$\fi}
\def \Msun {\ifmmode M_{\odot} \else $M_{\odot}$ \fi} 
\def \Lsun {\ifmmode L_{\odot} \else $L_{\odot}$ \fi} 
\def \Rsun {\ifmmode R_{\odot} \else $R_{\odot}$ \fi} 
\def \Msunpyr {\ifmmode M_{\odot}{\rm~yr}^{-1} \else $M_{\odot}{\rm~yr}^{-1}$ \fi} 
\def \hMsun {\ifmmode h^{-1}\,\rm M_{\odot} \else $h^{-1}\,\rm M_{\odot}$ \fi}

\def \LCDM {\ifmmode \Lambda{\rm CDM} \else $\Lambda{\rm CDM}$ \fi}
\def \sig8 {\ifmmode \sigma_8 \else $\sigma_8$ \fi} 
\def \OmegaM {\ifmmode \Omega_{\rm M} \else $\Omega_{\rm M}$ \fi} 
\def \OmegaL {\ifmmode \Omega_{\rm \Lambda} \else $\Omega_{\rm \Lambda}$\fi} 
\def \Deltavir {\ifmmode \Delta_{\rm vir} \else $\Delta_{\rm vir}$ \fi}
\def \rhocrit {\ifmmode \rho_{\rm crit} \else $\rho_{\rm crit}$ \fi}
\def \rhou {\ifmmode \rho_{\rm u} \else $\rho_{\rm u}$ \fi}
\def \zc {\ifmmode z_{\rm c} \else $z_{\rm c}$ \fi}

\def \rhos {\ifmmode \rho_{\rm s} \else $\rho_{\rm s}$ \fi} 
\def \rs {\ifmmode r_{\rm s} \else $r_{\rm s}$ \fi} 
\def \cvir {\ifmmode c_{\rm vir} \else $c_{\rm vir}$ \fi} 
\def \Rvir {\ifmmode r_{\rm vir} \else $R_{\rm vir}$ \fi}
\def \Vvir {\ifmmode V_{\rm vir} \else $V_{\rm vir}$ \fi} 
\def \Mvir {\ifmmode M_{\rm vir} \else $M_{\rm vir}$ \fi} 
\def \Nvir {\ifmmode N_{\rm vir} \else $N_{\rm vir}$ \fi} 
\def \Jvir {\ifmmode J_{\rm vir} \else $J_{\rm vir}$ \fi} 
\def \Evir {\ifmmode E_{\rm vir} \else $E_{\rm vir}$ \fi} 
\def \vvir {\ifmmode v_{\rm vir} \else $v_{\rm vir}$ \fi} 
\def \lam {\ifmmode \lambda \else $\lambda$ \fi} 
\def \lamp {\ifmmode \lambda^{\prime} \else $\lambda^{\prime}$ \fi} 
\def \Vmax {\ifmmode V_{\rm max} \else $V_{\rm max}$ \fi} 
\def \Mdm {\ifmmode M_{\rm dm} \else $M_{\rm dm}$\fi}

\def \Mgas {\ifmmode M_{\rm gas} \else $M_{\rm gas}$\fi} 
\def \Mcg {\ifmmode M_{\rm cg} \else $M_{\rm cg}$\fi} 
\def \Mhg {\ifmmode M_{\rm hg} \else $M_{\rm hg}$\fi} 
\def \Mdisc {\ifmmode M_{\rm disc} \else $M_{\rm disc}$\fi} 
\def \Md {\ifmmode M_{\rm d} \else $M_{\rm d}$ \fi} 
\def \Mda {\ifmmode M_{\rm d,0\%} \else $M_{\rm d,0\%}$\fi} 
\def \Mdb {\ifmmode M_{\rm d,20\%} \else $M_{\rm d,20\%}$\fi} 
\def \Mdc {\ifmmode M_{\rm d,40\%} \else $M_{\rm d,40\%}$\fi} 
\def \md {\ifmmode m_{\rm d} \else $m_{\rm d}$\fi} 
\def \Mb {\ifmmode M_{\rm b} \else $M_{\rm b}$\fi}
\def \Mbh {\ifmmode M_{\rm b,pri} \else $M_{\rm b,pri}$\fi} 
\def \Mbs {\ifmmode M_{\rm b,sat} \else $M_{\rm b,sat}$\fi} 
\def \zo {\ifmmode z_{0} \else $z_{0}$ \fi} 
\def \rd {\ifmmode r_{\rm d} \else $r_{\rm d}$\fi}
\def \rg {\ifmmode r_{\rm g} \else $r_{\rm g}$\fi}
\def \rb {\ifmmode r_{\rm b} \else $r_{\rm b}$\fi}
\def \rs {\ifmmode r_{\rm s} \else $r_{\rm s}$\fi}
\def \rc {\ifmmode r_{\rm c} \else $r_{\rm c}$\fi}
\def \rvir {\ifmmode r_{\rm vir} \else $r_{\rm vir}$\fi}
\def \rbh {\ifmmode r_{\rm b,pri} \else $r_{\rm b,pri}$ \fi} 
\def \rbs {\ifmmode r_{\rm b,sat} \else $r_{\rm b,sat}$ \fi}

\title[Star formation during mergers] 
{Star formation in mergers with cosmologically motivated initial conditions}

\author[W. Karman et al.] {Wouter
 Karman$^{1,2}$\thanks{karman@astro.rug.nl}, Andrea
 V. Macci\`o$^{2}$, Rahul Kannan$^{2,3}$, Benjamin
 P. Moster$^4$, \newauthor{Rachel S. Somerville$^{5}$}\\ 
\parbox[t]{\textwidth}{
$^1$ Kapteyn
 Astronomical Institute, University of Groningen, Postbus 800, 9700
 AV Groningen, the Netherlands\\ $^2$ Max-Planck-Institut f\"ur
 Astronomie, K\"onigstuhl 17, 69117 Heidelberg, Germany\\
$^3$ Department of Physics, Kavli Institute for Astrophysics and 
Space Research, Massachusetts Institute of Technology, Cambridge,
MA 02139, USA \\ $^4$
Kavli Institute for Cosmology, Institute of Astronomy,
University of Cambridge, Madingley Road, Cambridge CB3 0HA, UK\\ $^5$ Department of Physics and Astronomy,
 Rutgers University, 136 Frelinghuysen Road, Piscataway, NJ 08854, USA}
}

\begin{document} 

\date{\today}
  
\pagerange{\pageref{firstpage}--\pageref{lastpage}}\pubyear{2015}
 
\maketitle

\label{firstpage}

\begin{abstract}
We use semi-analytic models and cosmological merger trees to provide
the initial conditions for multi-merger numerical hydrodynamic
simulations, and exploit these simulations to explore the effect of
galaxy interaction and merging on star formation (SF). We compute
numerical realisations of twelve merger trees from $z=1.5$ to
$z=0$. We include the effects of the large hot gaseous halo around all
galaxies, following recent obervations and predictions of galaxy
formation models.  We find that including the hot gaseous halo has a
number of important effects.  Firstly, as expected, the star formation
rate on long timescales is increased due to cooling of the hot halo
and refuelling of the cold gas reservoir.  Secondly, we find that
interactions do not always increase the SF in the long term.  This is
partially due to the orbiting galaxies transferring gravitational
energy to the hot gaseous haloes and raising their temperature.
Finally we find that the relative size of the starburst, when
including the hot halo, is much smaller than previous studies showed.
Our simulations also show that the order and timing of interactions
are important for the evolution of a galaxy. When multiple galaxies
interact at the same time, the SF enhancement is less than when
galaxies interact in series.  All these effects show the importance of
including hot gas and cosmologically motivated merger trees in galaxy
evolution models.
\end{abstract}


\setcounter{footnote}{1}

\section{Introduction}
\label{sec:intro}

In the now well established model of Lambda Cold Dark Matter
($\Lambda$CDM) mergers and interactions between galaxies play an
important role in shaping their evolution
\citep[e.g][]{white1978}. Both observations and theoretical studies
have shown that interactions between galaxies can create tidal
bridges, streams and shells \citep[e.g.][]{zwicky1956,toomre1972,
  toomre1977,smith2007}. Minor mergers have been shown to thicken
disks and build up the spheroidal component
\citep{quinn1993,walker1996,villalobos2008,Villalobos2009,moster2010b,moster2011b,Zolotov2010}, while major mergers have been shown
to destroy disks and create elliptical and bulge dominated galaxies
(\citealt{hernquist1992,quinn1993,brook2004,cox2006,kazantzidis2008,kazantzidis2009}, 
but see \citealt{robertson2006b,governato2009,Querejeta2015} for disks surviving major mergers).

Mergers and interactions are very common --- over 70$\%$ of Milky Way-sized galaxies
have experienced a 10:1 merger since redshift $z \sim 1$ \citet[e.g.]{stewart2008,Lotz2011}. 
Furthermore, a significant amount of the mass formed in galaxies has
been assembled by accreting smaller galaxies, varying from 10 $\%$ to
$>$50$\%$ of the stellar mass for galaxies with $M_\star > 10^{10}
M_\odot$ \citep[e.g.][]{Oser2010,Lackner2012,Navarro2013}.
\citet{bell2006} used pairs at the same redshift in the COMBO-17
photometric redshift survey to study the rate of interactions between
galaxies. They found that in their sample $\sim 5\%$ of the galaxies
are in a close pair.  They also found that 50$\%$ of the massive
galaxies have undergone a major merger since $z\,=\,0.8$, and
therefore conclude that major mergers play a significant role in the
evolution of galaxies in this period.

An important effect of galaxy interactions and mergers is that they
cause enhanced star formation (SF)
\citep{larson1978,barton2003,lin2007,Hwang2011,Patton2011}, although 
weak in the local Universe \citep{Bergvall2003,Knapen2009}, and may trigger AGN activity
\citep[e.g.][]{sanders1988,treister2012, bessiere2014}. There is
increasing evidence that most extreme local star forming galaxies, the
Ultra Luminous InfraRed Galaxies (ULIRGS), are triggered by mergers
\citep[e.g.][]{sanders1996,borne1999,
  borne2000,bushouse2002,rupke2013a}, and an anticorrelation is found
between galaxy pair separation and the indicators for SF in the
optical
\citep[e.g. equivalent width of
  $H\alpha$,][]{sanders1988,barton2000,nikolic2004,Ellison2008,Patton2011}. 

Galaxy mergers have been studied extensively using idealised binary
merger simulations \citep[e.g.][]{toomre1977,hernquist1992,mihos1996,springel2005b,cox2008,hopkins2009a}.
In this approach, two
idealised (typically disk-dominated) galaxy models are created, each
within their own idealised dark matter halo, and these are set up on
an orbit which will lead to an eventual merger. The evolution during
the course of the merger is then computed using numerical N-body and
hydrodynamic techniques. These studies predict a strong link between
galaxy interactions and enhanced star formation.  During interactions,
the gravitational potential is disturbed, such that gas is able to
lose angular momentum and flow inwards. Because of this, the gas
density in the center increases rapidly, and a burst of SF (starburst)
takes place in the center of the galaxy
\citep{mihos1996,springel2005b,cox2008}.  \citet{cox2008} and
\citet{hopkins2009a} quantified this increased SF due to galaxy
interaction and found that it is proportional to the mass ratio of the
merging galaxies, as well as their cold gas fraction before the onset
of the merger. Many semi-analytic models use the fitting formulae
derived from these studies to model merger-driven starbursts
\citep[e.g.][]{bower2006,croton2006,delucia2007,monaco2007,somerville2008a}.

The majority of these previous idealised merger simulations include a
cold gaseous disk in the galaxy model, but did not consider the hot
gas component harboured in the dark matter halo.  Full cosmological
hydrodynamical simulations and semi-analytic models (SAMs) of galaxy
formation predict a large amount of hot gas around the galaxy, in
quasi static equilibrium with the gravitational potential
\citep[e.g.][but see also \citealt{cox2004} for the generation of a
  hot halo during mergers]{kauffmann1993,bower2006,somerville2008b,
  stinson2010,stinson2012, Brook2014}. Recently, several X-ray studies
found evidence for a hot gas component around the Milky Way
\citep{gupta2012,Kluck2013} and normal massive disc galaxies
\citep[e.g.][]{anderson2011,anderson2013} with temperatures of more
than $10^5$ K.

This hot gas component was first included in idealised hydrodynamical
merger simulations by \citet{Kim2009} and \citet{moster2011a}.  The
latter found that the cooling of the hot gaseous halo refuels the cold
gas in the center of galaxies, and as a result, SF can be sustained
for a longer time \citep[see also][]{choi2014}. Another interesting
result of \citet{moster2011a} is that the efficiency of mergers in
triggering SF is reduced in the presence of a hot halo.

An additional drawback of the idealised binary merger approach is that
initial conditions (such as the properties of the progenitor galaxies
and the orbit) must be specified \emph{a priori}, and will not
necessarily represent a cosmologically motivated distribution. In
addition, such studies have been almost entirely restricted to binary
mergers, while \emph{multiple mergers} (a sequence of more than two
mergers, in which the remnant does not have time to fully relax before
a subsequent merger) are a common phenomenon in the CDM universe
\citep[e.g.][MMS14]{Vaisanen2008,Duplancic2013}.

In this paper we take advantage of the method developed by
\citet{moster2014}, which combines cosmological merger trees and
semi-analytic models to specify well-motivated, cosmologically
representative initial conditions for a set of galaxy mergers, and
then carries out the evolution of the galaxies in the merger tree
using detailed numerical hydrodynamics. This approach is complementary
to fully cosmological hydrodynamic ``zoom-in'' simulations, in that
very high resolution can be achieved, and controlled experiments to
study specific physical processes can be carried out. This is
particularly relevant for testing and refining recipes for use in
semi-analytic models. We extend the work of \citet{moster2014} by
including the effect of supermassive black holes (BHs) and their
associated feedback in our merger simulations and studying the
modifications to the star formation rate (SFR) during mergers with
respect to models without BHs.

The paper is organised as follows: in Sec.~\ref{sec:methods} we
describe the methods that we employed in this work, including the code
that has been used, the physical recipes that we adopted and the
models that were used to create the initial conditions. In
Sec.~\ref{sec:sims} to \ref{sect:multi} we present our results. We
first compare simulations with and without the hot halo component in
Sec.~\ref{sec:sims}, then we inspect the effect of multiple mergers
in Sec.~\ref{sect:multi}, and finally we present the effects of BHs
in the simulations in Sec.~\ref{ss:BH}. We discuss the performance of
the code and possible shortcomings of the models in
Sec.~\ref{sec:discussion}, and then present our conclusions from this
study and possible improvements for future studies in
Sec.~\ref{sec:conclusion}.


\section{Methods}
\label{sec:methods}

\begin{figure}
\psfig{figure=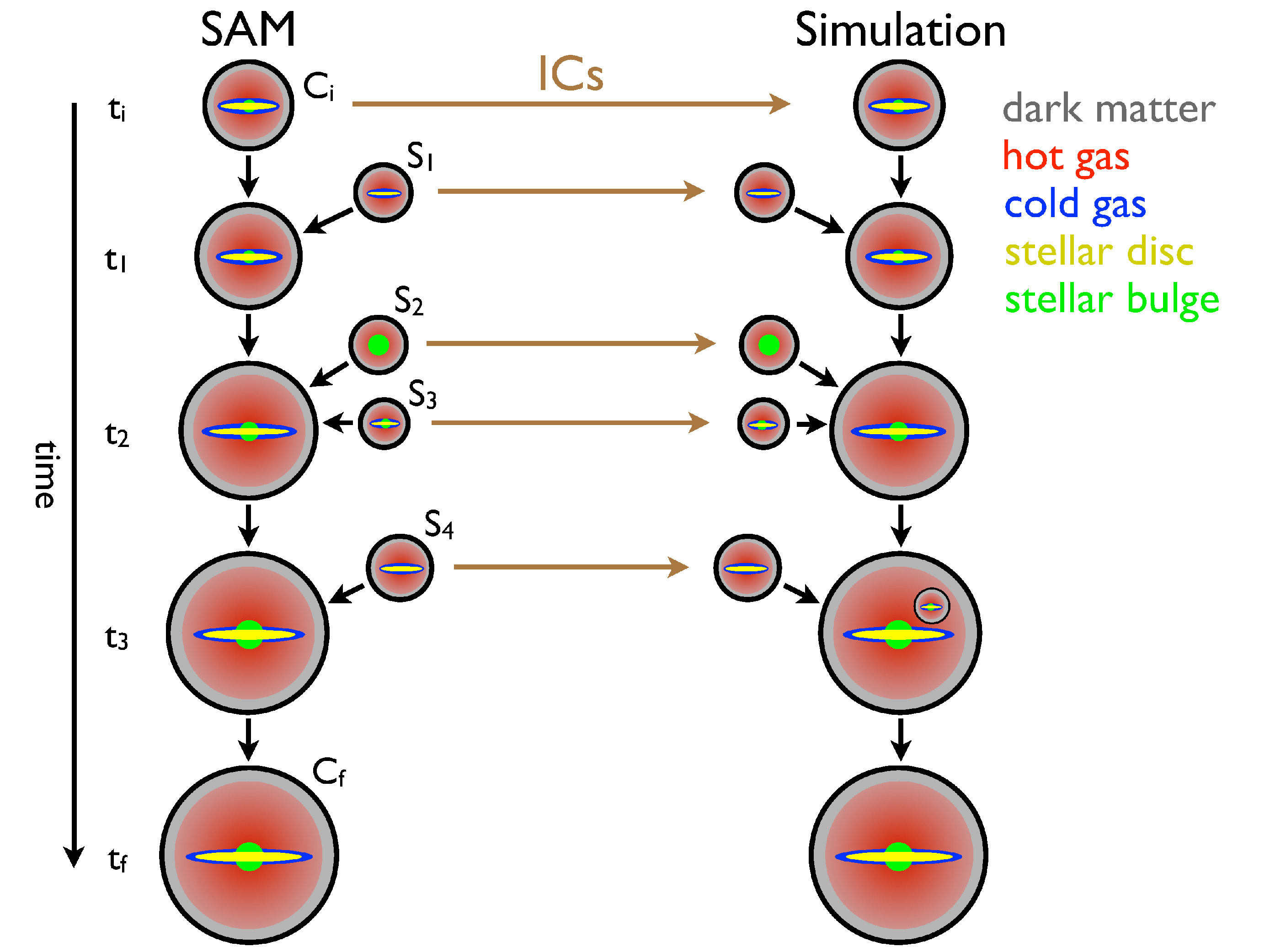,width=0.49\textwidth}
\caption{Schematic view of the scheme to combine SAMs and merger
  simulations (Fig. 6 from MMS14). On the left side the semi-analytic
  merger tree is shown with time running from top to bottom. At the
  starting time $t_i$, a particle realisation of the central galaxy is
  created using the properties of the central galaxy in the SAM at
  this time. This system is then simulated with the hydrodynamical
  code until the time of the first merger $t_1$, where a particle
  realisation of the first satellite $S_1$ galaxy is created, again
  using the properties predicted by the SAM, and introduced into the
  simulation with orbit properties taken from the cosmological
  $N$-body simulation. The resulting merger is simulated until the
  next galaxies ($S_2$ and $S_3$) enter the main halo at $t_2$, at
  which point they are also included in the simulation. This procedure
  is repeated for all mergers until the final time of the run $t_f$.}
\label{fig:samtosim}
\end{figure}

In this section we briefly describe the modelling approach used in this
paper, which is based on the simulated merger tree method described in
detail in MMS14. The method consists of three main steps: 1) merger
trees are extracted from a large dissipationless cosmological {\em
  N}-body simulation 2) a semi-analytic model is run in these merger
trees to ``populate'' the halos with galaxies and predict their
properties 3) each time two (or more) dark matter halos merge,
galaxy-halo models are created with properties predicted by the SAM,
placed on orbits specified by the cosmological simulation, and evolved
forward using a numerical hydrodynamic code. More details about each
of the three aspects are given below.

\subsection{Cosmological N-body simulation}
\label{sec:nbody}

The initial {\em N}-body simulation from which we extracted the merger
trees was performed with the {\sc gadget2} code (Springel 2005)
assuming WMAP3 cosmological parameters \citep{spergel2007} namely:
$\Omega_m=0.26$, $\Omega_{\Lambda}=0.74$,
$h=H_0/(100$~km~s$^{-1}$~Mpc$^{-1})=0.72$, $\sigma_8=0.77$ and
$n=0.95$. The simulation was done in a periodic box with a side length
of $100$ Mpc, and contains $512^3$ particles with a particle mass of
$2.8\times 10^8\Msun$ and a comoving force softening of $3.5$
kpc. Dark matter haloes were identified in the simulation snapshots
using a Friends of Friends (FOF) halo finder with a linking parameter
of $b=0.2$. Substructures inside the FOF groups are then identified
using the {\sc SUBFIND} code \citep{springel2001}.

\subsection{Hydrodynamical simulations}
\label{sec:hydro}

The hydrodynamical simulations in this work were performed using the
parallel TREE-SPH code {\sc pgadget3} last described by
\citet{springel2005a}. This code uses Smoothed Particle Hydrodynamics
\citep[SPH;][] {lucy1977,gingold1977,monaghan1992} to evolve the
gaseous medium using an entropy conserving scheme
\citep{springel2002}.

Important parameters in SPH simulations are the softening length
and the number of particles used. For the softening length $\epsilon$
we follow MMS14 and for every particles species we set it to
$\epsilon = \epsilon_1 \sqrt{m_{part}/10^{10}\Msun}$ \citep{dehnen2001},
where $\epsilon_1 = 32$ kpc for a particle of $10^{10}\Msun$.
For one of our typical simulations, the number of particles are 
$\sim 800000$ DM particles, $\sim 600000$ gas particles, and $\sim60000$
star particles, with masses of $\sim4\times 10^{5} \Msun$, 
$\sim3\times 10^{6} \Msun$, and $\sim2\times 10^{5} \Msun$ respectively.
In Table \ref{tab:partnrs} we give the number of particles for the major
components in our merger trees.

The multiphase interstellar medium is modelled following the SF recipe
introduced by \citet{springel2003}. This subgrid recipe stochastically
forms stars when cold gas clouds surpass a density threshold
($\rho_{\rm th}$). The newly formed stars eject a fraction of their
material, modelling Supernova (SN) driven winds, following
\citet{springel2003}.  The star formation is thus directly coupled to
SN feedback on a subgrid level, such that the wind mass is directly
proportional to the SFR, i.e. $\dot M_w= \eta \dot M_*$, where $\eta$
is the mass loading factor which quantifies the wind efficiency. These
gas particles receive a fixed fraction of the SN energy and are
decoupled from hydrodynamical interactions for a short period of time,
resulting in a constant initial wind speed $v_w$. 

The parameters for the multiphase ISM model are computed as outlined
in \citet{springel2003} in order to match the Kennicutt-Schmidt Law
\citep{kennicutt1998}. When a Kroupa IMF is adopted, the mass fraction
of massive stars is $\beta=0.16$ resulting in a cloud evaporation
parameter of $A_0=1250$ and a SN ``temperature'' of $T_{\rm
  SN}=1.25\times10^8{\rm~K}$. The SF timescale is set to
$t_*^0=3.5{\rm~Gyr}$ and for the galactic winds we adopt a
mass-loading factor of $\eta = 1$ and a wind speed of $v_w \sim 500
\kms$, which is an intermediate strength for winds.

\begin{table*}
 \centering
 \begin{minipage}{140mm}
 \caption
 {Summary of the parameters used for the simulations of merger trees
 and their fiducial values.} \null
 \begin{tabular}{@{}llr@{}}
 \hline Parameter & Description & Fiducial value\\ \hline
$z_i$ & Redshift at the start of the simulation & 1.5\\ $\mu_{\rm
 min}$ & Minimum dark-matter mass ratio & 0.0667/0.02\\ $\zeta$ &
 Ratio of scaleheight and scalelength of the stellar disc &
 0.15\\ $\chi$ & Ratio of scalelengths between gaseous and stellar
 disc & 1.5\\ $\xi$ & Ratio of gaseous halo core radius and dark
 matter halo scale radius & 0.22\\ $\beta_{\rm hg}$ & Slope parameter
 of gaseous halo & 0.67\\ $\alpha$ & Ratio of specific angular
 momentum between gaseous and dark halo& 4.0\\ $N_*$ & Expected final
 number of stellar particles in the central galaxy &
 $200\,000$\\ $\kappa$ & Ratio of dark matter and stellar particle
 mass & 15.0\\ $N_{\rm res,sat}$ & Ratio of satellite and central
 galaxy particle mass & 1.0\\ $N_{\rm min}$ & Minimum number of
 particles in one component & 100\\ $\epsilon_1$ & Softening length
 in kpc for particle of mass $m=10^{10}\Msun$ & 32.0\\ $t_0^*$ & Gas
 consumption time-scale in Gyr for SF model &
 3.5$^\dagger$\\ $A_0$ & Cloud evaporation parameter for SF model & 1250.0$^\dagger $\\ $\beta_{\rm SF}$ & Mass
 fraction of massive stars for SF model & 0.16$^\dagger
 $\\ $T_{\rm SN}$ & Effective supernova temperature in K for feedback
 model& $1.25\times10^{8\dagger}$\\ $\eta$ & Mass loading factor for
 wind model & 1.0\\ $v_{\rm wind}$ & Initial wind velocity in \kms
 for wind model & 500.0\\ $SeedRatio $& Ratio of seed mass and
 dynamical mass of BH particle & 0.01\\ \hline
 \null\\ \multicolumn{3}{l}{$^\dagger $ The SF parameters
 assume a Kroupa IMF.}
\label{t:smtparameters}
\end{tabular}
\end{minipage}
\end{table*}

Every simulation is repeated with BHs included and thermal AGN
feedback implemented as described by \citet{springel2005b}. BH
particles accrete mass by Eddington limited Bondi-Hoyle accretion
\citep{hoyle1939,bondi1944,bondi1952}, and this accretion is assumed
to deposit thermal energy into the surrounding medium at a rate
proportional to the accretion rate: $ \dot{E}_{\rm feed} \, = \,
\epsilon_{\rm f} \, \epsilon_{\rm r} \, \dot{M}_{\rm BH} \, c^2 $
where $ \epsilon_{\rm f}$ is the coupling strength of the feedback,
and is adopted from \citet{dimatteo2005}, and $ \epsilon_{\rm r}$ is
the standard conversion factor of rest-mass to energy for a standard
thin accretion disk from \citet{Shakura1973}. 

When galaxies merge, their BH are both brought to the center of the
remnant.  During this migration, AGN offset from the center and pairs
of AGN are predicted in the remnant. Observations have shown that
offset and dual AGN are present in a fraction of merging galaxies
\citep[e.g.][]{comerford2009,koss2012,comerford2014}.  Because the
movement and orbit of a second BH in galaxy mergers can be important
for the evolution of the remnant, we do not reposition the BH
particles on the local minimum of the potential.  On the other hand we
need to have stable BH particles, whose position is not affected by
the finite resolution of the simulation (for example due to two body
encounters with massive star particles).

In order to overcome this problem we defined a new parameter named
{\it SeedRatio} (SR), which defines the initial ratio between the seed
BH mass (i.e. the actual BH mass) and its dynamical mass.  The
dynamical mass is only used by the gravitational computations; by
increasing it with respect to the actual BH mass we ensure that two
body encounters between star (or gas) particles and the BH will not
substantially modify the BH trajectory. On the other hand all physical
processes associated with the BH (e.g. feedback) are governed by its
real mass and are not influenced by our choice of the initial SR
ratio.

It should be noted that as the BH gradually accretes mass at a subgrid
level, only the actual mass of the BH increases, while when another
particle is swallowed by the BH particle the dynamical mass increases.
We found that a ratio of 1000 for dynamical masses of BH particles to
gas particles ensures very stable BHs orbits in all our merger
simulations. At our resolution this results in $SR = 0.01$ (meaning
the BH dynamical mass is 100 times the BH mass).
\noindent
An overview of parameter values used in this work is given in Table
\ref{t:smtparameters}.

Before moving to the next section, we would like to comment on the
possible effects of the shortcomings of the {\sc pgadget3}
implementation of the SPH method on our results.  Recently,
\citet{Hayward2014} showed that the agreement between the adaptive
mesh refinement (AMR) approach and SPH methods is very good with
regard to SFRs in galaxy mergers.  They also show that the inclusion
of BHs does not significantly modify the small difference in the SFH
in SPH and AMR methods.  This shows that the conclusions in this paper
should be robust to the numerical treatment of hydrodynamics, in spite
of recent concerns about the accuracy of the classical SPH formulation
in certain situations \citep[e.g][]{Agertz2007}.

\subsection{Semi-analytic models and Simulated Merger trees}

In this paper we use the same approach as \citet{moster2011a} and
MMS14, and refer to those papers for a more extensive discussion.  In
our method, presented schematically in Fig.~\ref{fig:samtosim}, we
build a dark matter tree from a large-scale $N$-body simulation and
use the SAM of \citet{somerville2008a} to populate the $N$-body merger
tree with galaxies. The SAM predicts the baryonic properties for each
galaxy within the halo merger tree by applying simple but physically
motivated recipes for cooling and accretion, star formation and
chemical enrichment, stellar feedback, black hole growth and AGN
feedback, and morphological transformation and starbursts in
mergers. The radial sizes of disks are modelled using an angular
momentum based approach which has been shown to predict a size-mass
relation that is in very good agreement with observations up to $z\sim
2$ \citep{somerville2008b}. These SAMs have been extensively tested
and shown to reproduce many fundamental statistical properties of
galaxies, including stellar mass functions and luminosity functions
from $z\sim 0$--6, disk gas fractions, and the fraction of different
morphological types from $z\sim 0$--3
\citep{somerville2008a,somerville2012,Porter2014,Brennan2015}.

In Fig. \ref{fig:samtosim} we show an example of such a merger tree,
where time runs from top to bottom. At a chosen time $t_i$ we extract
a halo with its baryonic properties and merger tree. In this simple
example the main system experiences four mergers, at $t_1$, $t_2$
(two) and $t_3$. At $t_i$ we obtain the properties from the SAM, and
create a particle realisation of the main halo ($C_i$), represented by
the brown arrow.  In Fig. \ref{fig:samtosim}, this generated galaxy is
shown in the top right, and consists of a dark matter halo (grey), a
hot gaseous halo (red), a cold gaseous disc (blue), a stellar disc
(yellow) and a stellar bulge (green). We then simulate the evolution
of this galaxy up to $t_1$, at which point we add a particle
realisation of satellite galaxy $S_1$ to the simulations.  While we
obtain all baryonic properties from the SAM, the orbital parameters of
$S_1$ are directly taken from the dark matter $N$-body simulation. In
this way we create a cosmologically motivated merger history.

After the inclusion of this satellite galaxy, we evolve the system in
our simuations up to $t_2$. At $t_2$ we add two more satellites ($S_2$
and $S_3$) at the virial radius using the same method as before and
continue the evolution of the system. We repeat this procedure up to
the end of the simulation at $t_f$. By using this method we naturally
include multiple mergers, such as at $t_2$ and at $t_3$ where one of
the satellites is still orbiting when satellite $S_4$ enters the halo.
The growth of the main halo occurs through two channels. First, the
main halo accretes gas and dark matter, which takes place from $t_i$
to $t_1$ or from $t_3$ to $t_f$ in the given example. Second, smaller
satellite galaxies feed the main halo through mergers, e.g. from $t_1$
to $t_2$.

\subsubsection{Galaxy Initial conditions}

We follow the description provided by \citet{springel2005b}, with the
extension of \citet{moster2011a} to create particle realisations for 
our initial conditions. Each galaxy consists of two stellar components (a disc 
and a bulge with masses \Mdisc and \Mb), two gaseous components (a cold 
disc and a hot halo with masses \Mcg and \Mhg), and a dark matter halo 
with mass \Mdm.

The stellar and gaseous discs are set up such that they are
rotationally supported and have surface density profiles that decline
exponentially outwards.  The parameter $\chi$ couples the scale length
of the cold gaseous disc to that of the stellar disc by $\rg =
\chi\rd$, and the vertical structure of the discs is determined by a
radially independent sech$^2$ profile with scale height $z_0$. The
velocity dispersion in the stellar disc is set equal in the vertical
and radial direction, while the gas temperature is determined by the
equation of state (EOS). A balance of the pressure due to the EOS and
the gravitational potential then determines the vertical structure of
the disc selfconsistently.

For the stellar bulge and dark matter halo, we assume a
\citet{Hernquist1990} profile with scale lengths $r_{\rm b}$ and
$r_s$. The dark matter halo is characterised by a concentration
parameter $c=\rvir/\rs$ and a halo spin $\lambda$. We assume that the
stellar bulge is spherical and non-rotating.

\citet{moster2011a} calibrated the properties for a hot gaseous halo
in order to match the cosmic star formation rate. We adopt a spherical
slowly-rotating density profile, that follows the observationally
motivated $\beta$-profile \citep[e.g.][]{cavaliere1976,
  eke1998}. Requiring that pressure supports an isotropic model and
hydrostatic equilibrium within the gravitational potential fixes the
temperature profile of the hot gaseous halo. A further calibration
needed was the specific angular momentum of the gaseous halo, and we
follow \citet{moster2011a} in setting the specific angular momentum of
the gaseous halo $j_{\rm hg}$ to a multiple $\alpha$ of the
dark-matter specific angular momentum $j_{\rm dm}$, such that $j_{\rm
  hg} = \alpha j_{\rm dm}$, with $\alpha=4$.

In simulations with BHs, the initial mass of the BHs is extracted from
the SAM, but due to resolution limitations we only include BHs with a
mass larger than $10^6\Msun$.  All the stellar and gaseous parameters
(e.g. \Mdisc, \Mb, \Mcg, \Mhg, \rd, etc.)  are taken from the SAM
predictions, while the dark matter parameters (\Mdm, c, $\lambda$) and
all the orbital parameters (position and velocity) of the mergers are
taken directly from the {\em N}-body simulation.

\subsection{Merger tree selection}

We set the lower mass limit of satellite galaxies to be included in
the numerical realisation to 1/15 of the mass of the central
galaxy. We chose this limit because it has been previously shown that
mergers with a mass ratio $\frac{M_{host}}{M_{sat}}>5$ barely influence the SFR
\citep{cox2008}\footnote{We did run three merger trees with a cut of
  1/50 to validate this assumption and found no substantial
  differences in the SF history.}. Although we could therefore exclude
any satellites with $\frac{M_{host}}{M_{sat}}>5$ from our simulations,
we choose to be conservative and set the limit at 
$\frac{M_{host}}{M_{sat}}=15$, in order to catch any possible 
secondary effects on the merger trees of slightly smaller satellites.

From the {\em N}-body+SAM simulation we selected 12 merger trees based
on the following criteria:
\begin{itemize}
 \item The final mass of the halo has to be of the same order of
 magnitude as that of the Milky Way (e.g. $2-3\times 10^{12}\Msun$)
 \item The merger tree must contain at least one major merger
 \item The total sample has to contain mergers of every mass ratio
from 1 to 5 in steps of 0.5.
 \item The galaxy with the smallest mass ratio should enter the virial
 radius after redshift 1.5
\end{itemize}

Milky Way mass galaxies sit at the peak of SF efficiency
(Fig. \ref{fig:SHM}), and the transition between disc to spheroid
dominated systems also occurs at this mass scale
\citep{Conselice2006}. Both of these effects are strongly related with
mergers. In addition, the steep relation between halo and stellar mass
at lower masses results in a drop of the influence of major stellar
mass mergers for lower mass galaxies. At higher mass, it has been
shown that additional feedback recipes are needed to prevent
overcooling
\citep[e.g.][]{Vernaleo2006,Gabor2011,Vogelsberger2014,Schaye2015}.

We invoke the second and fourth criteria to satisfy the aims of our
study, i.e. determine the effect of the merger ratio on the star
formation history in cosmologically motivated mergers at $z<1.5$. The
third criterion ensures that we sample the full range of merger
ratios.

The final merging process of the two most massive galaxies is not
disturbed significantly by other galaxies in the majority of the
trees, so during the final merger they are similar to binary
mergers. However, we selected three merger trees that harbor multiple
simultaneous mergers, in order to study the importance of three body
interactions \citep[e.g.][]{moreno2012}.

We also evolved the central galaxy in isolation for the same time
interval, such that we have a control sample without mergers. We will
refer to these simulations as the ``isolated'' cases.  Each merger
tree is simulated from a starting redshift $z=1.5$ down to $z=0$.
Finally we reran all merger trees with the addition of BHs and AGN
feedback.  All together we have a suite of nearly 100 different
galaxies and simulations, leading to a substantial statistical sample.

\section{Star formation evolution}
\label{sec:sims}

We computed the stellar-to-halo-mass ratio for all of our initial and
final central galaxies, and compared it to the constraints derived by
\citet{moster2013} using abundance matching.  This comparison shows
that the initial conditions and final properties of our galaxies are
consistent with these constraints.

\begin{figure*}
\centering
\begin{minipage}{180mm}
\psfig{figure=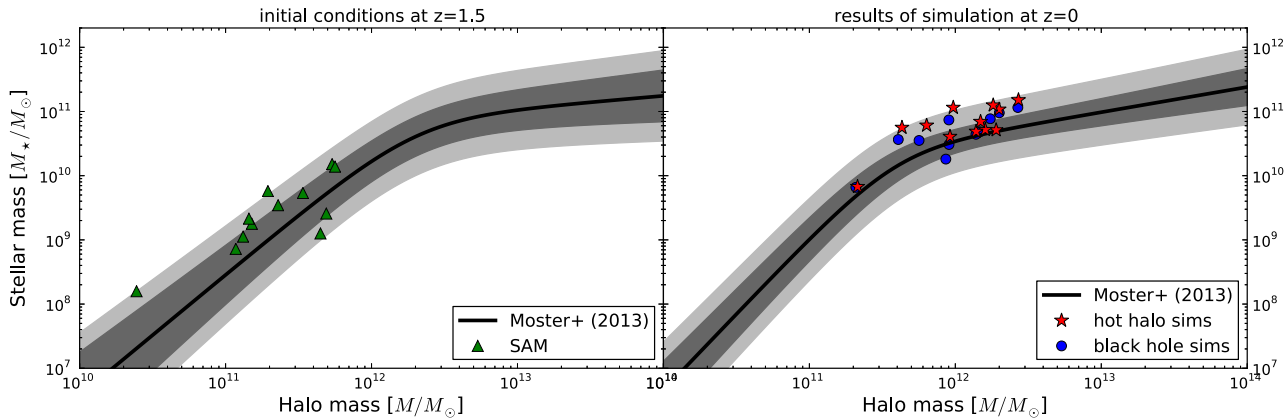,width=0.98\textwidth}
\caption{Stellar mass versus dark-matter mass within the virial
  radius. The left plot shows the initial galaxy properties obtained
  from the SAM, while the right plot shows the resulting simulated
  central galaxies at redshift zero. The red stars show the
  simulations with the hot halo model and the blue circles show the
  simulations with the hot halo model and BHs.  Overplotted are the
  abundance matching constraints derived by \citet{moster2013}. The
  dark shaded area shows the 1$\sigma$ range on the constraints, and
  the light shaded area is the 2$\sigma$ range. \label{fig:SHM}}
\end{minipage}
\end{figure*}

\begin{table}
 \begin{tabular}{l|l|l|l|l|l}
 \bf Tree nr &$M_{host}$&$M_{sat}$& $\mu$ &$M_{\star , host}$&$M_{\star , sat}$\\
  &$10^{10}\Msun$ & $10^{10}\Msun$& &$10^{10}\Msun$& $10^{10}\Msun$ \\ \hline
 661ws1 & 12.7 & 2.04 & 6.23 & 0.385 & 0.0895\\ 
661ws2 & 24.7 & 8.39 & 2.95 & 2.18 & 0.955 \\ 
661bh1 & 13.9 & 2.11 & 6.59 & 0.313 & 0.226\\ 
661bh2 & 22.7 & 7.85 & 2.89 & 1.81&  1.05 \\ 
872ws & 27.4 & 9.01 & 3.04 & 4.72 & 2.95 \\ 
872bh & 23.2 & 6.47 & 3.59 & 3.86& 2.01\\ 
990ws & 21.6 & 8.98 & 2.41 &2.85 & 2.10 \\ 
990bh & 19.6 & 8.80 & 2.22 &2.42 & 2.13 \\ 
1166ws1a & 3.21 & 0.77 & 4.17 & 0.781 & 0.104\\ 
1166ws1b & 3.21 & 1.93 & 1.66 & 0.781 & 0.577\\ 
1166bh1a & 3.46 & 1.05 & 3.30 & 0.650& 0.132\\ 
1166bh1b & 3.46 & 2.13 & 1.62 & 0.650& 0.494\\ 
1178ws & 7.42 & 2.66 & 2.79 & 0.700 & 0.452  \\ 
1178bh & 7.46 & 2.88 & 2.59 &0.657&0.320\\ 
1188ws1 & 5.14 & 3.77 & 1.36 & 0.914 & 0.508\\ 
1188ws2 & 9.68 & 1.27 & 7.62 & 2.83& 0.365\\ 
1188ws3 & 11.7 & 7.41 & 1.58 & 3.63& 1.66\\ 
1188bh1 & 4.85 & 3.49 & 1.39 &0.825&0.487\\ 
1188bh2 & 7.72 & 0.948 & 8.14 &2.25&0.364\\ 
1188bh3 & 9.15 & 4.07 & 2.25 &2.77&6.60\\ 
1415ws1a & 1.04 & 1.19 & 0.87 & 0.197 & 0.221 \\ 
1415ws1b & 1.04 & 1.50 & 0.69 & 0.197& 0.269\\ 
1415ws2 & 7.13 & 4.71 & 1.51 & 1.95 & 1.05\\
 1684ws & 10.8 & 5.33 & 2.02 &2.78&0.605\\
 1684bh & 11.1 & 4.79 & 2.32 &2.81&0.567\\ 
2096ws & 4.49 & 3.95 & 1.14 & 0.728 & 0.373\\ 
2096bh & 4.36 & 3.34 & 1.31 &0.675&0.352\\ 
3747ws1 & 4.11 & 0.605 & 6.79 & 0.182& 0.0414\\ 
3747ws2 & 3.97 & 2.82 & 1.41 & 1.39 & 0.366 \\ 
3747bh1 & 3.72 & 0.856 & 4.35 &0.167&0.106\\ 
3747bh2 & 3.83 & 2.34 & 1.64 &1.28&0.199
 \end{tabular}
\caption{Properties of the progenitor galaxies immediately before they
  merge. Col. (1): the id number of the merger tree, (2) total mass of
  host galaxy within $R_{\rm s,host}$, (3) total mass of satellite
  galaxy within $R_{\rm s,sat}$, (4) merger ratio $\mu =
  \frac{M_{host}}{M_{sat}}$, (5) stellar mass of host within $R_{\rm
    s,host}$, (6) stellar mass of satellite within $R_{\rm s,sat}$. In
  Col. 1, the label {\em ws} means that the simulation was run with
  only the hot halo model, while the {\em bh} label means a simulation
  with hot halo and BHs. Numbers after these labels give the number of
  the merging event, and letters a subset of multiple merger
  events. Because merger trees 780 and 2809 do not have any mergers by
  $z=0$, they are not included in this
  table.\label{table:mergerratio}}
\end{table}

The results are shown in Fig. \ref{fig:SHM}; although the initial
conditions are mostly located within 1$\sigma$ of the relation, the
galaxies at redshift $z = 0$ mostly lie above the 1$\sigma$
region. This is not unexpected because we selected merger trees
with active merger histories, which will have preferentially larger
amounts of mass growth and SF than the average.  Merger trees run with
BHs do agree better with the stellar-halo mass relation, as feedback
from AGNs reduces the SF.  Overall our galaxies do not seem to suffer
from a serious overcooling problem and are consistent with
observations.

We simulated a total of twelve merger trees from redshift $z=1.5$ to
$z=0$, and summarise the characteristics of all mergers in Table
\ref{table:mergerratio}. The merger ratio $\mu =
\frac{M_{host}}{M_{sat}}$ is an important parameter in this
study. Satellite galaxies, after entering the halo of the central
galaxy and before merging with it, lose mass due to dynamical friction
and ram pressure stripping \citep[e.g.][]{boylankolchin2008,
  weinmann2012, Chang2013}.  This mass evolution introduces some
arbitrariness in the definition of the merger ratio, because it
depends on when it is computed. We decided to compute it right before
the merger between the galaxies, in order to take into account the
mass loss. We first fitted the dark matter profile of both galaxies,
before the merger, to an NFW profile \citep{navarro1997} and obtained
a value for the scale radius $r_s$. We then calculated the total mass
(dark matter+gas+stars) inside $r_s$ at the time of merging ($t_1$)
for both satellite and central, and used these masses to define $\mu$.

The mass ratios obtained with this method are in very good agreement
with results from substructure finders like {\sc subfind}
\citep{springel2001}.  This method was always able to distinguish
between the central and satellite galaxies at very close encounters,
whereas this proved difficult for {\sc subfind}.

\subsection{The effect of the Hot Halo}
\label{ss:hothalo}

\begin{figure*}
\centering
\psfig{figure=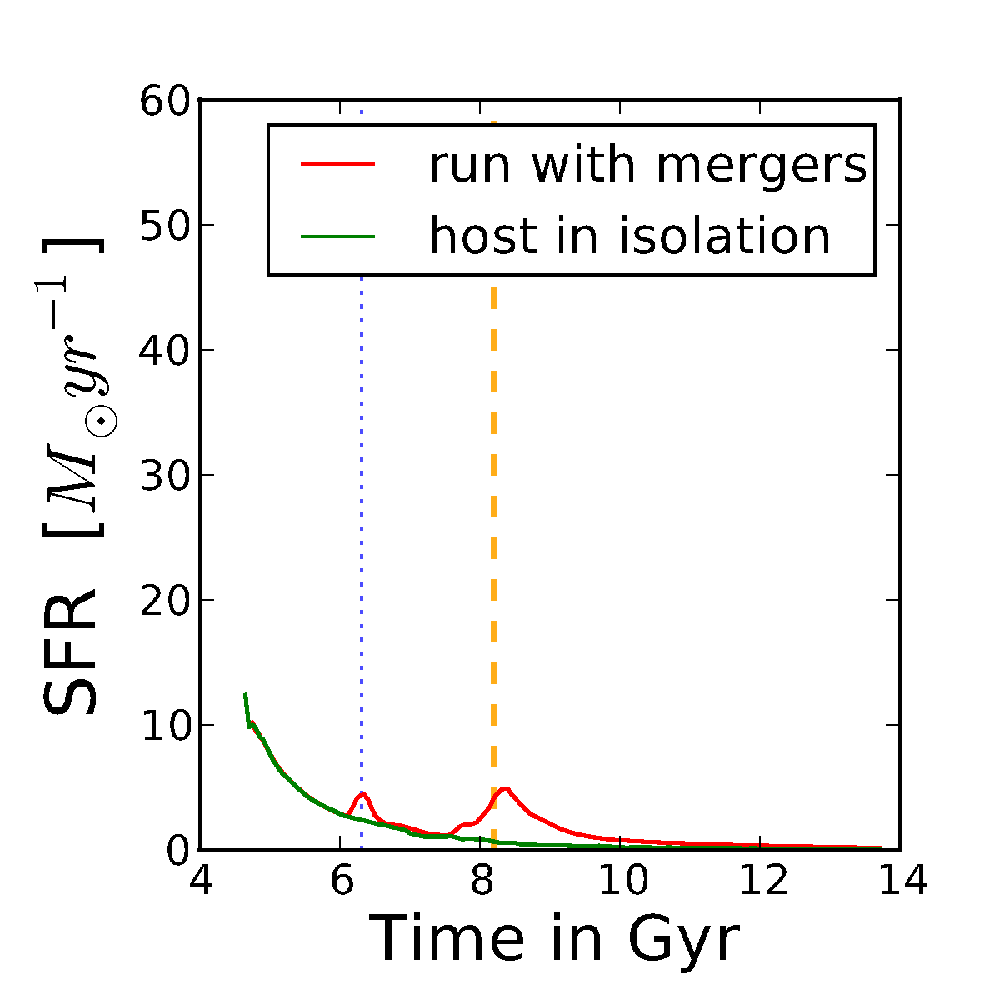,width=0.39\textwidth}
\psfig{figure=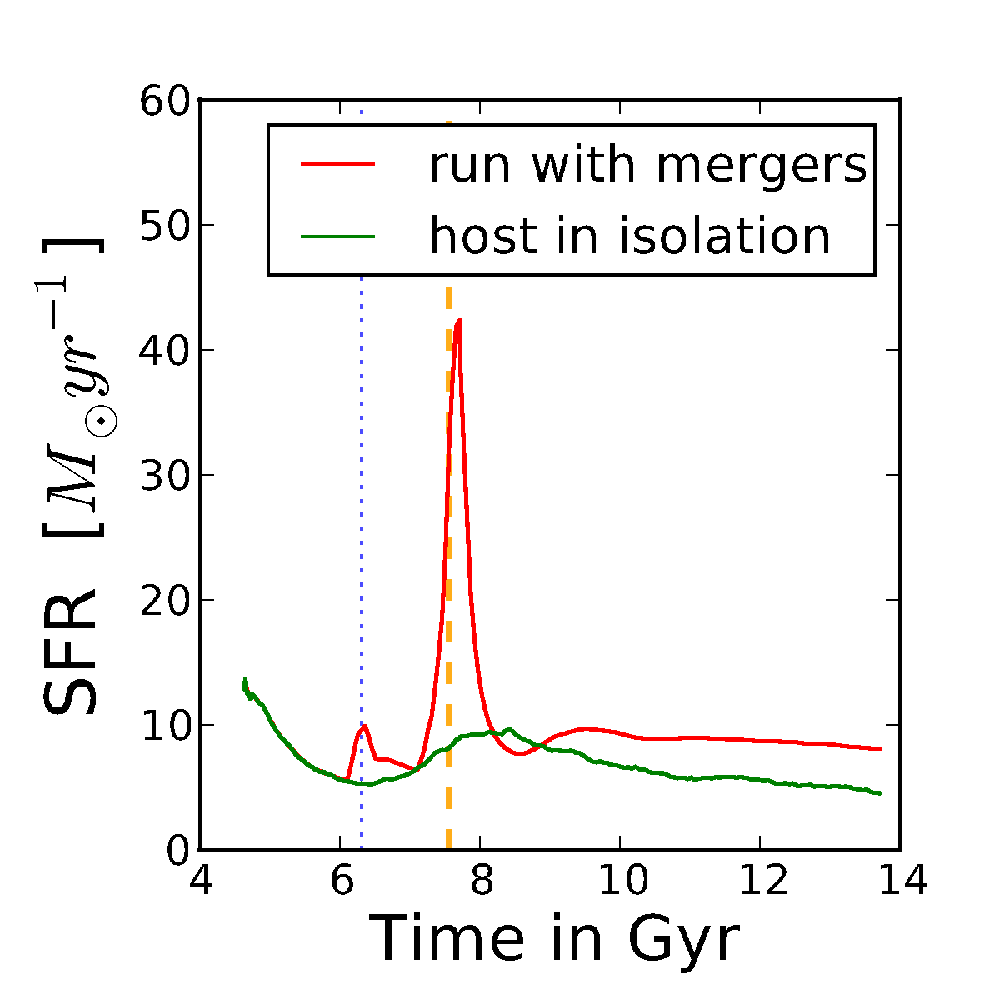,width=0.39\textwidth}
\caption{The SFR of merger tree 990. The left panel shows the
  simulations where no hot halo is present, and the right panel shows
  the simulations with a hot gaseous halo around the galaxy. The red
  line shows the SFR for the primary galaxy with the simulated merger,
  and the green line is the primary galaxy simulated in isolation. The
  vertical dashed orange line indicates the time when the merger
  occurs  while the blue dotted line shows flyby's. \label{fig:SFR990}}
\end{figure*}

As a first test, we compare simulated merger trees with and without
the inclusion of the (observationally expected) hot gaseous halo
component.  We present the SFR for merger tree 990 in Fig
\ref{fig:SFR990}, for a run without hot halo (left panel) and with hot
halo (right panel). In each panel the results of the merger simulation
are shown in red, while the isolated run (e.g. central galaxy without
satellites) is shown in green. It is important to note that the
merger ratio of two galaxies can vary after the removal of the
hot halo, due to a different fraction of hot gas.

The cold-gas-only simulation is in agreement with previous studies
\citep[e.g.][]{cox2008}; the SFR drops rapidly when the gas is depleted,
and a starburst occurs during the final merging process. The starburst
increases the SFR by an order of magnitude and there is almost no SF
after the merger, because all the gas has been consumed during the
starburst.  The quenching of SF due to starvation is particularly
clear in the isolated run, where we see a progressive
fading of SF as time goes by. We caution that the relative increase
in SF can be less than an order of a magnitude, for example,
\citet{Dimatteo2007,Dimatteo2008} find an average increase of 
a factor of $\sim3$ during mergers.

The simulation including a hot halo (right panel) shows a quite
different behavior \citep[as already noted by][]{moster2011a}.  The
hot halo cools and therefore the cold gas in the center of the galaxy
is refuelled and SF can be sustained for the whole duration fo the
simulation; this is particularly clear in the isolated run.

Similar to MMS14, we see an important difference between the
simulations with and without a hot halo during the merger.  The
starburst occurring in the simulation with a hot halo produces many
more stars than the one in the simulations without a hot halo,
however, the relative increase with respect to the isolated galaxies
is smaller. In the simulations without a hot halo we observe an order
of magnitude difference between the isolated and merging galaxies,
which is significantly larger than the factor of $\sim 6$ difference
in the simulations with a hot halo.  This difference is mostly due to
the different evolution of the SFR in the isolated runs.  The isolated
galaxy without the hot halo depletes a large fraction of its gas by
forming stars in the first 2 Gyrs (before the merger happens), but
when it has a hot halo surrounding it, it is able to keep forming
stars for a longer period (because it replenishes its cold gas
reservoir). This reduces the relative efficiency of a merger in
triggering additional SF.

\begin{figure*}
\centering
\psfig{figure=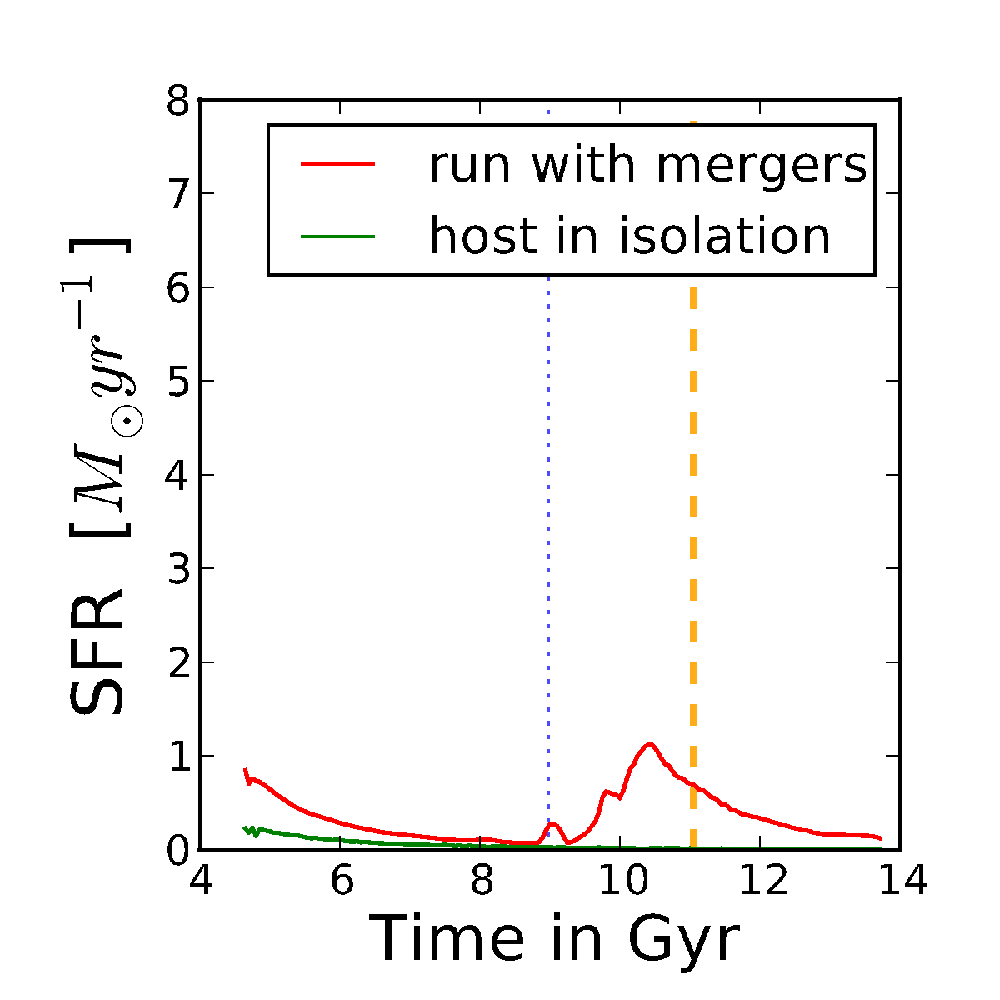,width=0.39\textwidth}
\psfig{figure=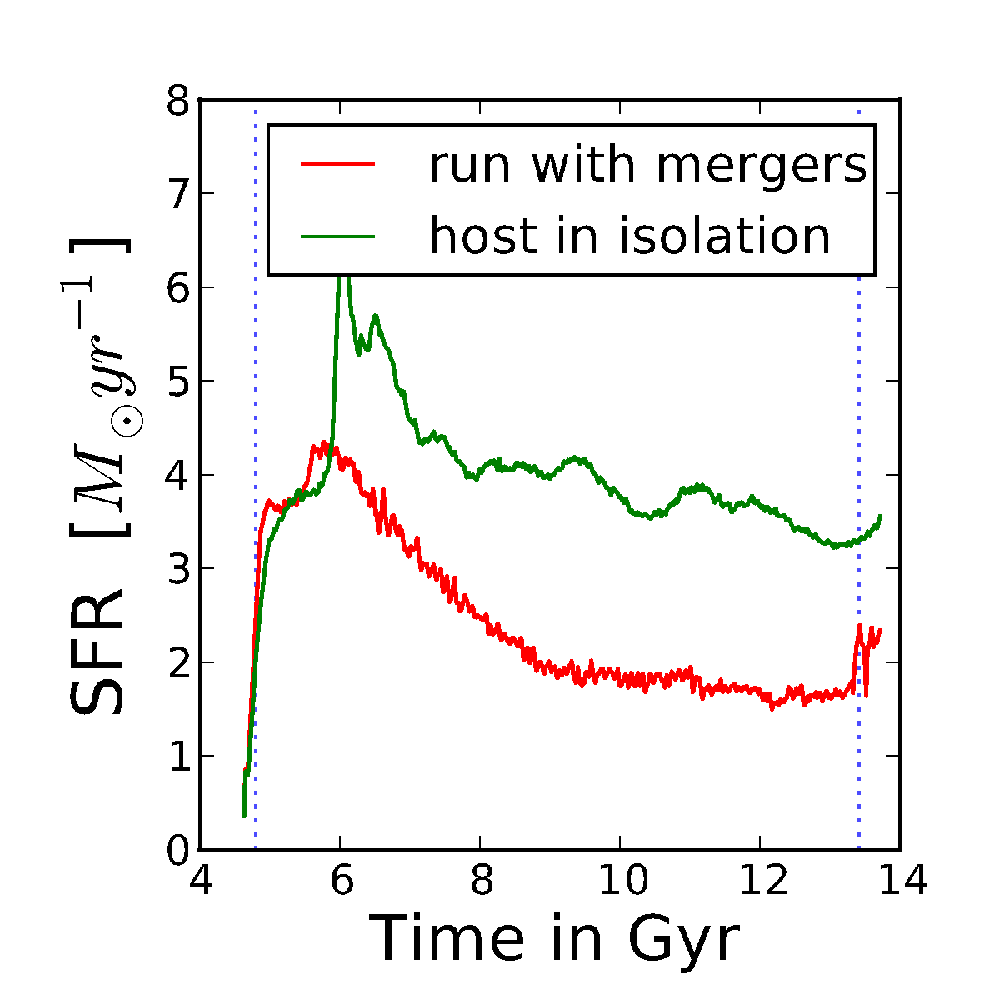,width=0.39\textwidth}
\caption{Same as Fig. \ref{fig:SFR990} but for merger tree number 780.\label{fig:SFR780}}
\end{figure*}

In Fig. \ref{fig:SFR780} the SF History (SFH) of merger tree 780 is
shown with and without the hot halo component.  For the simulation
without a hot halo, the behavior is very similar to tree 990, with an
increased SF at $t\sim10$ Gyr. This ``bump'' in the SF clearly shows
the effect of the merger, but it is not present in the simulation with
a hot halo even though the mass ratio is very similar. 
When looking directly at the simulation snapshots at $z =
0$, we see indeed that the two galaxies did not merge and are 
at a distance of 73 kpc, despite two
close passages (with distances of $\sim$100 kpc and $<$5 kpc) and in
contrast to expectations based on the dark matter only simulation.
\footnote{Since in this merger tree, and also in merger tree 2809, the
  two galaxies do not merge by $z=0$, we did not use these merger
  trees in the remainder of the paper.}

Even more interesting is the {\it decrease} of SF in the merging hot
halo simulation compared to the isolated galaxy. This decrease is seen
over the full simulated period. This result is in contrast to the
findings of previous merger simulations
\citep[e.g.][]{cox2008,hopkins2009a}, where the SFR only experienced a
positive boost due to interactions of galaxies.  In order to better
understand the origin of this SFR reduction we looked in more detail
into the cold/hot gas fraction in the merger simulation.

\citet{sinha2009} and \citet{moster2011a} already found that when two
galaxies merge the hot halo component experiences heating due to the
creation of shocks and the dissipation of the satellite galaxy orbital
energy.  This results in a net increase of the gas temperature and in
a longer cooling time.

\begin{figure*}
\psfig{figure=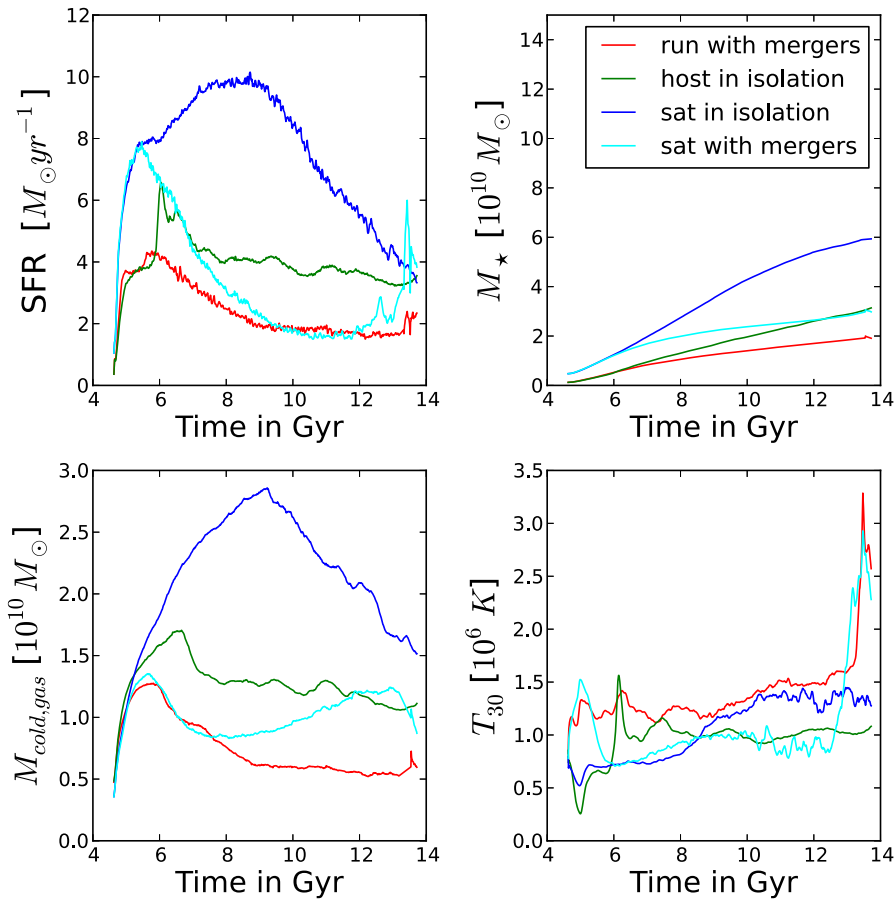,width=0.7\textwidth}
\caption{Merger tree 780: the top left plot shows the evolution of the
  SFR of this merger tree from redshift 1.5 to 0; the top right panel
  is the stellar mass over this period; the bottom left shows the
  amount of cold gas present in the galaxy, and the bottom right panel
  is the average temperature of the gas between 30 and 40 kpc around
  the galaxy. The red (green) line always shows the primary
  (secondary) galaxy in the merger simulation, while the blue (cyan)
  lines shows quantities for the primary (secondary) galaxy simulated
  in isolation. \label{fig:Temp780}}
\end{figure*}

In Fig. \ref{fig:Temp780} (bottom right panel) we show the temperature
for the halo gas at a distance from 30 to 40 kpc from the galaxy
center.  This temperature is shown for the central galaxy in isolation
(green) during the merger (red) and for the incoming satellite galaxy
again in isolation (blue) and in the merger simulation (cyan).  For
both galaxies (central and satellites) the gas temperature is higher
in the merger run than in isolation, confirming the results of
\citet{moster2011a}.  This higher gas temperature increases the
cooling time, and reduces the amount of cold gas available for SF, as
shown in the bottom left panel of Fig.  \ref{fig:Temp780}.  For the
satellite galaxies the effects of ram pressure and tidal
stripping make the difference between the isolated and merger cases
even more pronounced.  Finally the increased temperature of the hot
gas and the subsequent reduced amount of cold gas decreases the SFR
and produces a lower final stellar mass (right and left upper panels
in Fig. \ref{fig:Temp780}) in the merger run with respect to the isolated
case, making the merger an overall ``inhibitor'' for SF.

The increase of the SFR in the isolated host galaxy at $\sim6$ Gyr 
is caused by disk instabilities that originate from the accretion
of cooling gas. Because the cooling rate is lower in the merger, 
the amount of accreted gas is insufficient to introduce the instabilities
observed in the isolated case after 6 
Gyr\footnote{The distance between the two galaxies increases from 200 kpc at
5.5 Gyr to 500 kpc at 7.5 Gyr, showing that during this period there
is no significant interaction.}.

Another important effect that might be at play in this galaxy is 
tidal stripping. If the first encounter at 4.8 Gyr generates strong tidal effects, 
a significant fraction of the cold gas can be expelled from the galaxies. 
This stripping would make the gas unavailable for star formation, and
provides another method of reducing star formation that proposed above. 
However, when visually inspecting this galaxy,
we find no signs of strong tidal effects, such as tails, during the first
encounter. This shows that the difference in star formation between the
isolated and merging galaxies is not dominated by tidal effects.

When looking at the complete suite of 12 merger trees, we see a number
of important trends. First, in all simulations the inclusion of the
hot halo component increases the overall amount of SF both in the
isolated runs and during mergers.  The hot halo is able to refuel the
cold gas reservoir and prevent the artificial quenching of SF due to
gas depletion.  Second, starbursts of varying sizes always occur
during close encounters between galaxies. Mergers are able to funnel
gas into the central region of the galaxy, where it is rapidly
converted into stars.  Most starbursts show a factor of $\sim 6$
increase in SFR with respect to the isolated run of the same galaxy
(see next section for a more quantitative results).  Finally we found
that mergers can also decrease the total SFR especially over long
periods. Orbiting satellites increase the temperature of hot halo gas,
by both the creation of shocks and transfer of orbital energy into
thermal and kinetic energy of the gas through dynamical friction and
hydrodynamic drag.  The increased temperature of the hot halo leads to
longer cooling times, diminishing the amount of hot gas able to cool
and reducing the overall amount of SF.

\begin{table*}
 \begin{tabular}{l|l|l|l|l|l|l|l|l|l}
 \bf Tree nr & $\bf t_{\rm 1}$ & $\bf t_{\rm 2}$ & $\bf
 \mu$& f$_{\rm gas}$ & f$_{\rm bulge, host}$ & f$_{\rm bulge, sat}$ &$\bf M_{\rm burst}$&$\bf M_{\rm cold}$ & $e_{\rm burst}$ \\ 
 & {\it Gyr} & {\it Gyr} & & & & & $10^{10} \Msun$ & $10^{10} \Msun$ &\\ 
\hline 
 661  & 5.23  & 6.34 & 6.23 & 0.65 & 0.001 & 0.16 & 0.232  & 1.933 & 0.12\\
 661  & 7.87  & 9.54 & 2.95 & 0.32 & 0.20 & 0.001 & 0.135  & 3.631 & 0.037\\ 
 872  & 9.78  & 10.51& 3.04 & 0.10 & 0.23 & 0.56 & 0.171  & 2.645 & 0.064\\
 990  & 7.26  & 8.47 & 2.41 & 0.24 & 0.19 & 0.18 & 1.079  & 2.056 & 0.52\\
 1166 & 7.60  & 8.30 & 4.17 & 0.50 & 0.22 & 0.33 & 0.123  & 1.161 & 0.11\\
 1166 & 8.30  & 8.85 & 1.66 & 0.26 & 0.43 & 0.11 & 0.0365 & 1.444 & 0.025\\
 1166$^\dagger$ & 6.97  & 8.52 & 1.12 & 0.52 & 0.005 & 0.008 & 0.490 & 1.407 & 0.35\\
 1178 & 7.80  & 9.80 & 2.79 & 0.52 & 0.047 & 0.18 & 0.538  & 2.854 & 0.19\\
 1188 & 6.38  & 7.59 & 1.36 & 0.44 & 0.061 & 0.079 & 0.708  & 1.923 & 0.37\\
 1188$^\dagger$ & 6.41 &7.36& 1.52 & 0.44 & 0.10 & 0.10 & 0.736 & 1.430&0.51\\
 1188 & 10.27 & 10.76 & 7.62 & 0.22 & 0.24 & 0.15 & 0.0617& 1.141 & 0.054\\
 1188 & 12.85 & 13.72 & 1.58 & 0.16 & 0.42 & 0.14 & 0.172 & 1.923 & 0.089\\
 1415$^{\dagger\dagger}$ & 7.70 & 8.85 & 0.87& 0.55 & 0.26 & 0.08 & 0.671 & 1.139 & 0.59\\
 1415$^{\dagger\dagger}$ & 7.70 & 8.85 & 0.69& 0.55 & 0.26 & 0.11 & 0.671 & 1.139 & 0.59\\
 1415 & 11.65 & 13.00 & 1.51 & 0.18 & 0.45 & 0.10 & 0.458 & 0.804 & 0.57\\
 1684 & 8.33  & 11.08 & 2.02 & 0.33 & 0.15 & 0.002 & 1.012 & 3.120 & 0.32\\
 2096 & 7.81  & 10.03 & 1.14 & 0.44 & 0.055 & 0.024 & 2.125 & 2.838 & 0.75\\
 3747 & 5.25  & 6.45  & 6.79 & 0.81 & 0.009 & 0.013 & 0.247 & 1.525 & 0.16\\
 3747 & 9.95  & 11.40 & 1.41 & 0.37 & 0.21 & 0.20 & 0.813 & 3.058 & 0.27
\end{tabular}
\caption{Results of the hot halo only simulations, where only ten out 
of the initial twelve merger trees contained major mergers. The first column
 gives the tree number, the second and third the begin and end of the
 starburst, the fourth the mass ratio from Table
 \ref{table:mergerratio}. In column five, six, and seven the cold-gas fraction within
$R_{\rm s}$ of both galaxies, the bulge fraction of the host, and the bulge fraction of the
satellite are given. The eighth column is the burst mass from equation
 \ref{calc:Mburst}, and the ninth is the amount of cold gas in the
 central region. The burst efficiency as defined in equation
 \ref{def:bursteff} is given in the last column. ($^\dagger$): The
 least massive of the three simultaneously interacting galaxies 
is removed and the merger is reran, see Sec.
 \ref{sect:multi}; ($^{\dagger\dagger}$): Combination of three
 galaxies, unable to distinguish contributions, so both merger ratios
 are given, see also Sec.
 \ref{sect:multi}. \label{table:effwinds}}
\end{table*}

\subsection{Starburst efficiency with hot halo}

In MMS14 it was shown that star formation enhancement in multiple
mergers can be significantly different than in a series of binary
mergers. Here we extend the work of MMS14 on this issue. In this work
we quantify our results in a more useful and descriptive way, so as to
better compare the star formation during mergers in isolated binary
mergers and in multiple mergers as predicted by cosmological merger
trees. Following the approach of \citet{cox2008} and
\citet{moster2011a}, we introduce the starburst efficiency parameter
$e_{\rm burst}$, defined as the fraction of cold gas present in the
pre-merger galaxy that is consumed by the additional burst of SF.

\begin{equation}
 e_{\rm burst} = \frac{M_{\rm burst}}{M_{\rm premerger,cold,gas}}\, .
\label{def:bursteff}
\end{equation}
Where the burst mass $M_{\rm burst}$ is defined as the additional amount of
stellar mass formed due to the merger between $t_1$ and $t_2$, which defines
the beginning and the end of the burst (e.g. $t_1=7.5$ and $t_2=8.5$ in the 
left panel of Fig. \ref{fig:SFR990}, and should not be confused with $t_1$ 
and $t_2$ from Fig. \ref{fig:samtosim}, which are the entry times of different
satellites into the simulation):

\begin{equation}
 M_{\rm burst} = \int_{t_1}^{t_2} SFR(t) {\rm dt} - \frac{1}{2}
 \left[SFR\left({\rm t_1}\right) + SFR\left({\rm t_2}\right)\right]
 ({\rm t_2} - {\rm t_1})\, ,
\label{calc:Mburst}
\end{equation}

\begin{figure} 
\includegraphics[width=0.49\textwidth]{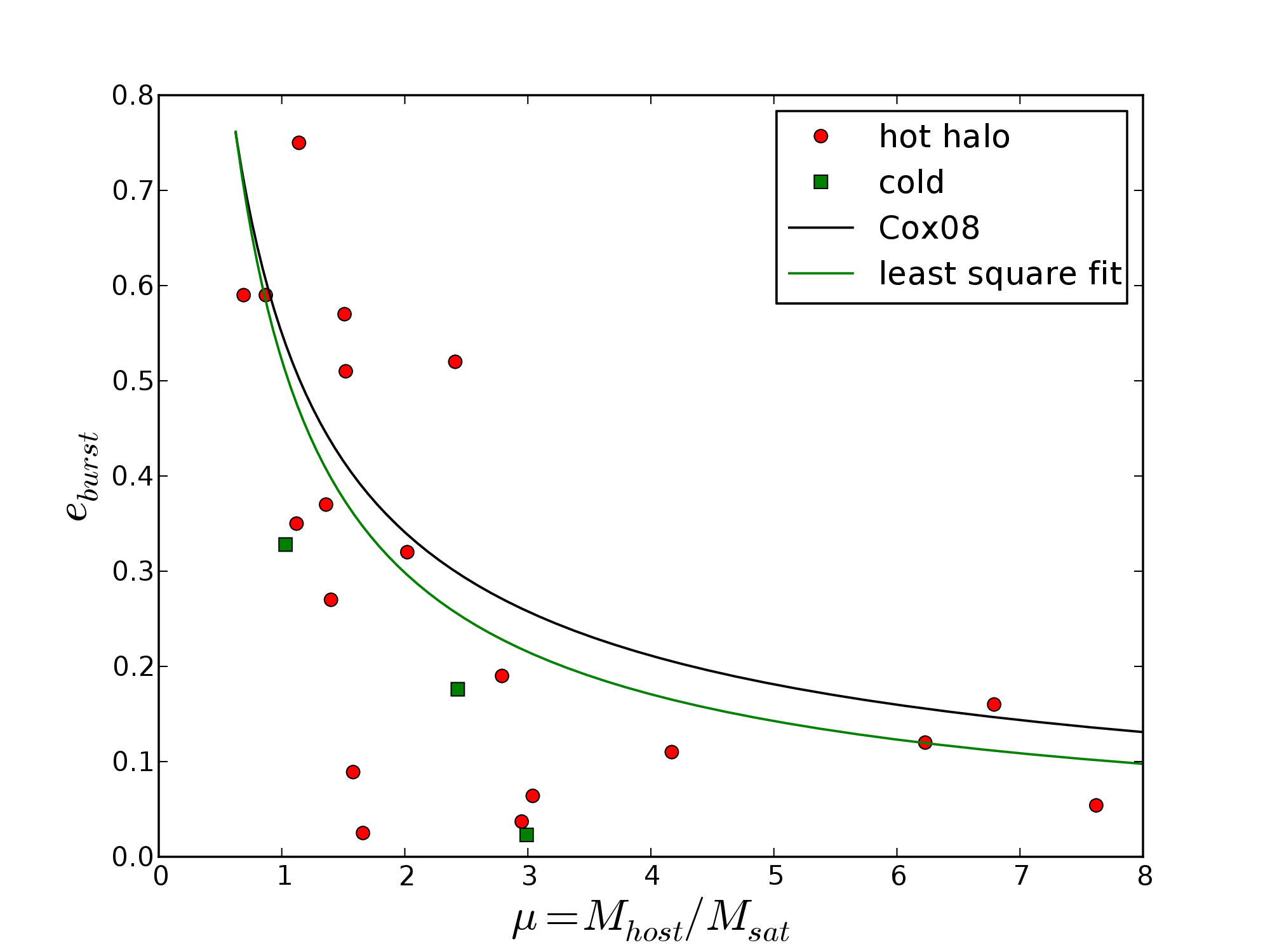}
\caption{Mass ratio versus starburst efficiency. The red points are
  the mergers in the simulations with a hot halo, while the green squares
show the simulations without a hot halo. The black line is
  the starburst efficiency relation of \citet{cox2008} and the green
  line represents the least squares fit to our
  results.  \label{fig:bursteff} }
\end{figure}

\noindent
The burst masses and burst efficiencies from our suite of mergers are
listed in Table \ref{table:effwinds}. In addition, this table shows
the bulge fraction $f_{\rm bulge}$, the gas fraction $f_{\rm gas}$, and other important
properties of our mergers. The bulge fraction is calculated
as described in detail by Kannan et al. (in prep.), but is based
on a decomposition of the stellar component in a bulge and a disk,
using the angular moment of the stellar particles, similarly to the approach
of \citet{Scannapieco2011}.  
Fig. \ref{fig:bursteff} shows the relation
between the burst efficiency and the galaxy merger ratio $\mu$. For
$\mu < 5$ there is a large scatter in the values of $e_{\rm burst}$,
while for $\mu>5$ the starburst efficiency seems to converge to values
around 0.15 (though the scatter is difficult to assess due to small
number statistics).

Following \citet{cox2008}, we performed a power law fit to the burst
efficiency as a function of mass ratio:
\begin{equation}
 e_{\rm burst} = e_{1:1} \left(\frac{M_{\rm sat}}{M_{\rm host}}
 \right)^\gamma .
\label{eqn:bursteff2x}
\end{equation}
We find that for our simulations including the hot halo the best fit
is found for $e_{1:1} = 0.52$ and $\gamma = 0.81$, and the resulting
relation is shown in Fig. \ref{fig:bursteff}, along with the fit with
the original values proposed by \citet[][Eq. 5]{cox2008}.

The values for $e_{1:1} $ and $\gamma$ are comparable to those found
by \citet{cox2008}, but the large scatter in the $e_{\rm burst}$
values suggests that other parameters besides the mass ratio are
impacting the burst efficiency. The larger scatter is due to three
factors. First, the number of mergers included in this study is larger
than in \citet{cox2008}, who fit to ten data points. Second, we use a
range of cosmologically based orbits and initial galaxy properties,
whereas \citet{cox2008} only varied the mass ratio. \citet{cox2008}
show that the gas fraction, bulge-to-disk ratio and pericentric
distance also influence the starburst efficiency, however, this is
insufficient to explain the large scatter. This is most clearly
demonstrated by comparing merger trees 990 and 1178, which have
similar properties, but significantly different burst
efficiencies. Third and possibly most important, the more complex
setting of the full merger tree influences the state of the galaxies.

\begin{figure} 
\includegraphics[width=0.49\textwidth]{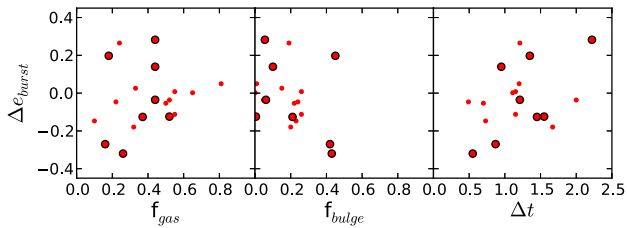}
\caption{Difference between the burst efficiency compared to the power
  law fit to our results shown in Fig.~\ref{fig:bursteff}, $\Delta
  e_{\rm burst} = e_{\rm burst} - e_{\rm burst, PL}$. The difference
  is shown as a function of the gas fraction $f_{\rm gas}$, the bulge
  fraction of the main galaxy $f_{\rm bulge}$, and the duration of the
  starburst $\Delta t = t_2 - t_1$, which is a tracer of the orbital
  properties. The large circled points are all major mergers with
  $1<\mu<2$, while the smaller points are all more minor
  mergers. \label{fig:bursteff_ws_scat} }
\end{figure}

In Fig. \ref{fig:bursteff_ws_scat} we examine the origin of the
scatter in more detail, and we show that the departure from the fitted
power law cannot be explained by a single parameter. The scatter
around the power law shows only weak trends with $f_{\rm gas}$, or
$f_{\rm bulge}$. We use the duration of strong interaction, $\Delta
t$, as a tracer of orbital influence, rather than trying to describe
the orbit by an ellipse. If one did describe the orbit with an
ellipse, the parameters of the ellipse would change between every time
step due to interactions with other galaxies. Therefore, the initial
orbital parameters are a very poor tracer of the final orbital
parameters \citep[see also][]{Tsatsi2015}.  The final orbital
parameters are better traced by $\Delta t$, where a more tangential
orbit has a higher value of $\Delta t$. Although the overall
distribution of the difference in burst efficiency seems random, there
might be a trace of a positive trend with increasing $\Delta t$ when
we only select mergers with a similar merger ratio ($1<\mu<2$, large
points in Fig. \ref{fig:bursteff_ws_scat}). The positive correlation between
the duration of the encounter and the strength of the starburst was also found
 by \citet{Dimatteo2007}. In summary, this shows
that many factors contribute to the efficiency of the merger-triggered
starburst and it may be difficult to devise a simple fitting
formula. We will show in Sec. \ref{ss:BH} that this conclusion holds
after the addition of BH feedback to the simulations.

\begin{table}
 \begin{tabular}{l|l|l|l|l|l|l|l|l|l}
 \bf Tree & $\bf t_{\rm 1}$ & $\bf t_{\rm 2}$ & $\bf
 \mu$ &$\bf M_{\rm burst}$&$\bf M_{\rm cold}$ & $e_{\rm burst}$ \\ 
 & {\it Gyr} & {\it Gyr} & & $10^{10} \Msun$ & $10^{10} \Msun$ &\\ 
\hline 
 780  & 9.35  & 11.65 & 1.03 & 0.111  & 0.340 & 0.328 \\
 872  & 11.00  & 12.60 & 2.99 & 0.008  & 0.341 & 0.023 \\
 990  & 7.55 & 9.70 & 2.43 &  0.187  & 1.066 & 0.176 \\
\end{tabular}
\caption{Results of the simulations without a hot halo. 
Only three
out of twelve merger trees were simulated with these initial conditions.
The columns correspond to columns 1, 2, 3, 4, 8, 9, and 10 from table
 \ref{table:effwinds}. \label{table:cold_effwinds}}
\end{table}

\section{Multiple mergers}
\label{sect:multi}

\begin{figure*}
\includegraphics[width=0.7\textwidth]{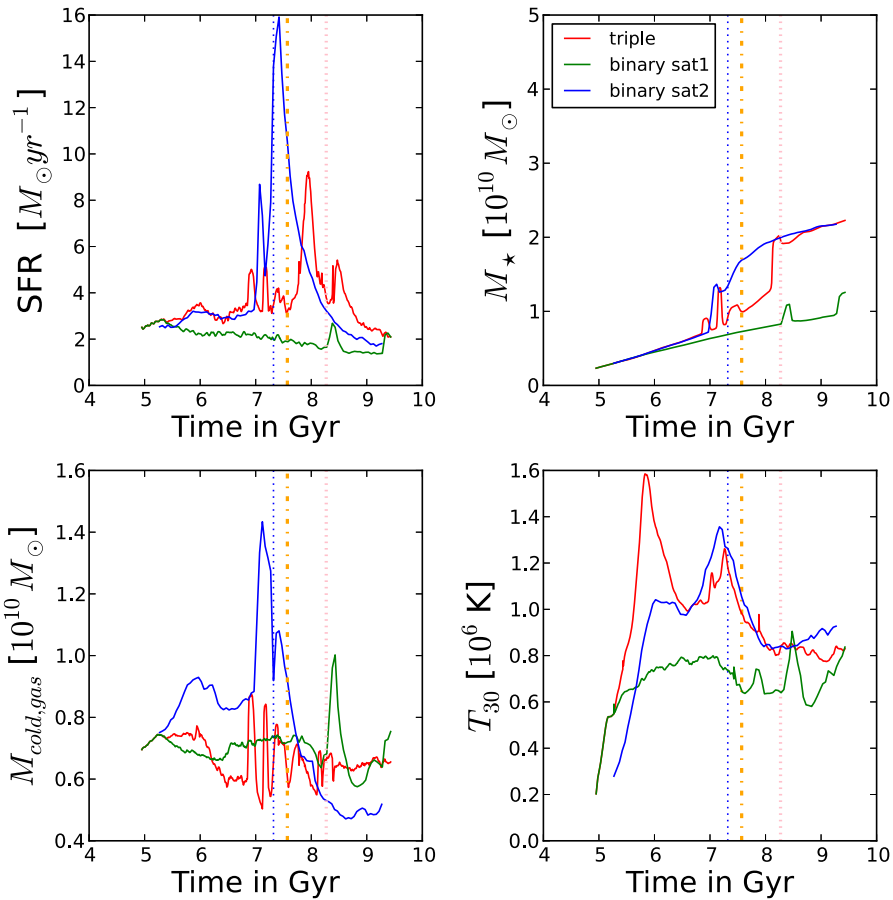}
\caption{The multiple mergers in tree 1166. The top left plot shows
  the SFR, the top right plot the total stellar mass, and the bottom
  left plot the cold-gas mass in the central volume with radius of 20
  kpc. The bottom right plot shows the average temperature of the halo
  gas between 30 and 40 kpc ($T_{30}$). The red line is the central
  galaxy in the original simulation, and the green line shows the
  simulation where only the central galaxy and first satellite is
  included. The blue line shows the simulation where the first
  satellite is left out, and only the two most massive galaxies are
  included (the central galaxy and SAT2).  The vertical orange
  dash-dotted line and pink dashed line indicate the time when the two
  satellites finally merge with the host galaxy ($\mu=4.17$ and
  $\mu=1.66$) in the simulation with a triple merger, while the
  vertical dotted blue line indicates the time when the galaxies
  merged in the binary merger simulation including only the most
  massive galaxies ($\mu=1.12$). \label{fig:multi1}}
\end{figure*}

In addition to situations where only two galaxies interact
simultaneously, so-called binary mergers, it also happens that more
than two galaxies interact with each other or share a common halo.
These multiple merger events are common in the LCDM universe
\citep[e.g., see][for an observed merging triplet]{Vaisanen2008}, and
can have important consequences for galaxy evolution.  In more than
50\% of all merger pairs since $z<5$, a second satellite enters the
common halo within five dynamical times after the first satellite
entrance (MMS14).

Nearly all previous studies on the effect of major mergers on SF
focused only on binary mergers. Here we use our cosmologically
motivated merger tree to study the effects of multiple
strongly-interacting mergers on SF.  We selected three merger trees
(namely 1166, 1188, 1415) where at least three galaxies strongly
interact with each other during a merger.  We refer to these multiple
strong interactions as multiple merger events.

In Fig. \ref{fig:multi1} we present the results for tree 1166.  In
each panel three different simulations are shown: 1) the multiple
merger where the central galaxy first merges with the first satellite
(SAT1) and then the second satellite (SAT2) enters the virial radius
before the end of the first merger; 2) the binary merger between the
central galaxy and the first satellite (without SAT2); 3) a binary
merger between the central galaxy and SAT2, which happens to be more
massive than SAT1.

\begin{figure*} 
\includegraphics[width=\textwidth]{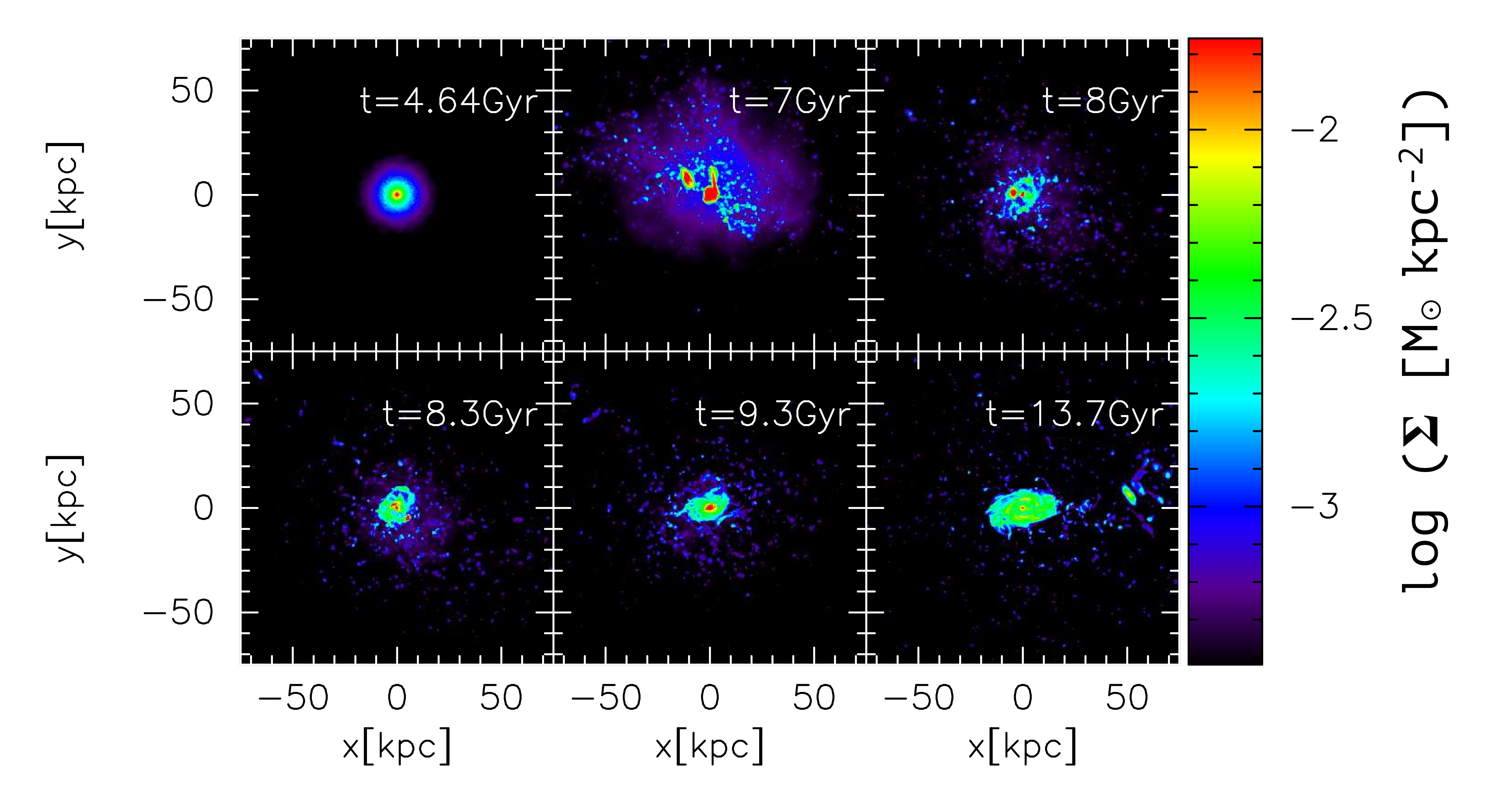}
\caption{Distribution of gas in tree 1166. The different snapshots
  represent the start of the simulation (t=4.64 Gyr), before, during,
  and after the multiple merger (t=7, 8, 8.3, 9.3 Gyr), and the end of
  the simulation (t=13.7 Gyr). Different colours correspond to
  different gas surface densities as shown in the
  key. \label{fig:gas1166} }
\end{figure*}

The effect on SF is stronger in the merger of the two more massive
galaxies (central and SAT2) than in the case of the triple merger. In
the binary merger the starburst is more pronounced and more
extended. The first satellite has a very minor impact on the SFR,
moreover in the case of a triple merger SAT1 has an overall negative
effect on the SFR, which is reduced with respect to the binary merger
between central and SAT1.

Although the stellar mass formed in the starburst of the merger with
only SAT2 is larger than in the original simulation, the stellar mass
that resides in the galaxy at the end of both simulations is very
similar. The difference in stellar mass formed is compensated by the
accretion of the stellar mass native to SAT1, and therefore results in
similar total stellar masses.

The reduced SF in the multiple merger is clearly due to the reduced
amount of cold gas, as shown by the third panel in
Fig. \ref{fig:multi1}. In the multiple merger simulation the orbiting
of the third galaxy increases the temperature of the hot gas (see the
fourth panel of Fig. \ref{fig:multi1}) delaying the cooling
process. In Fig. \ref{fig:gas1166} we show the distribution of gas in
this merger tree at important stages of this interaction.

\begin{figure*}
\includegraphics[width=0.9\textwidth]{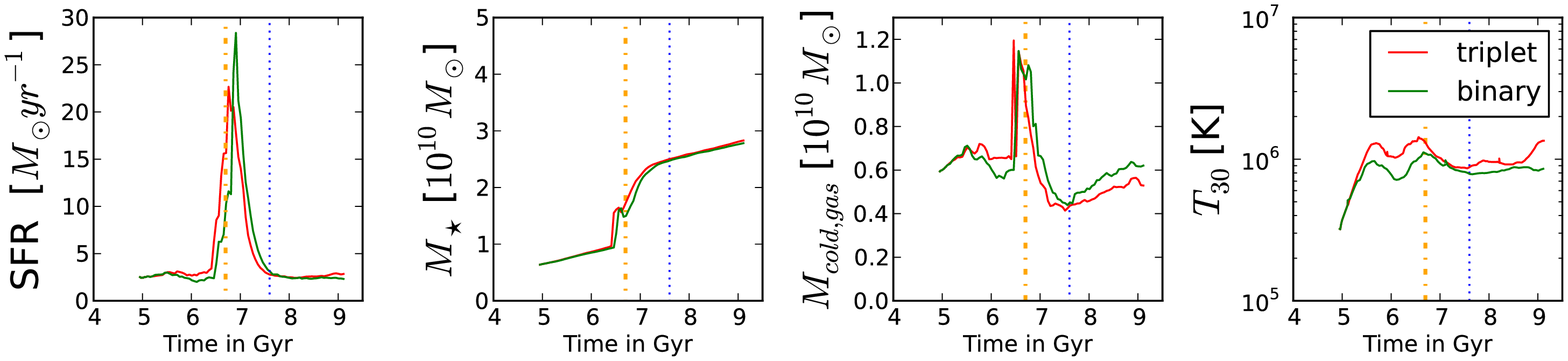}
\includegraphics[width=0.9\textwidth]{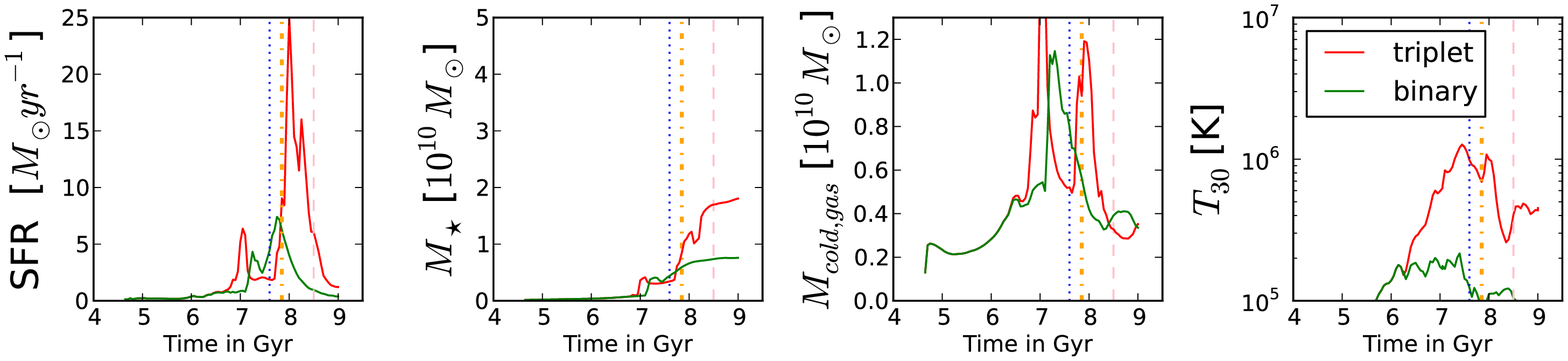}
\caption{The multiple mergers in trees 1188 and 1415. The leftmost
  columns show the SFR, the second column the total stellar mass, and
  the third column the cold-gas mass within a radius of 20 kpc. The
  rightmost column shows the average temperature $T_{30}$. The red
  line is the central galaxy in the original simulation, and the green
  line shows the simulation where only the central galaxy and first
  satellite is included.  The vertical orange dash-dotted and pink
  dashed lines (only in tree 1415) indicate the time when the two
  satellites merge with the host galaxy ($\mu=1.36$ for 1188, and
  $\mu=0.87$ and $\mu=0.69$ in 1415) in the simulation with a triple
  merger, while the vertical dotted blue line corresponds to the time
  when the galaxies merged in the binary merger where only the most
  massive galaxies are included in the simulation ($\mu=1.52$ for
  1188, and $\mu=0.87$ for 1415). \label{fig:multi2}}
\end{figure*}

Fig. \ref{fig:multi2} shows the same panels as Fig. \ref{fig:multi1}
for merger trees 1188 (upper row) and 1415 (lower row). In these plots
we see similar results. In 1188 SAT1 is the most massive and indeed
produces a larger starburst than the triple merger, in agreement to
what we found in 1166.  In merger tree 1415 we encountered a more
complex situation. From Table \ref{table:mergerratio} it can be seen
that the merger ratio between the central galaxy and the two
satellites SAT1 and SAT2 is less than 1 using our definition of
$\mu$.  This means that both the satellites are more massive than the
central galaxy, and that the merger between the two satellites will be
triggering the most SF, also in the triple merger.  Simply replacing
the central galaxy by one of the satellites and keeping the same
orbital parameters, does not result in a merger before $z=0$. The
orbital parameters would therefore have to be significantly adjusted
in order to have a merging binary system of the two satellites, and
this would complicate the comparison.  For all these reasons it is
very difficult to compare the results of 1415 with the other multiple
mergers.

\section{Effect of black hole feedback}
\label{ss:BH}

Feedback from accretion supermassive black holes can have an important
effect in regulating SF during mergers
\citep{dimatteo2005,hopkins2006,Choi2012}.  During mergers, gas can be
efficiently funnelled towards the center of the galaxy triggering
rapid accretion onto the BH (Quasar phase) and a subsequent
substantial deposition of energy into the interstellar medium.

\begin{figure*} 
\psfig{figure=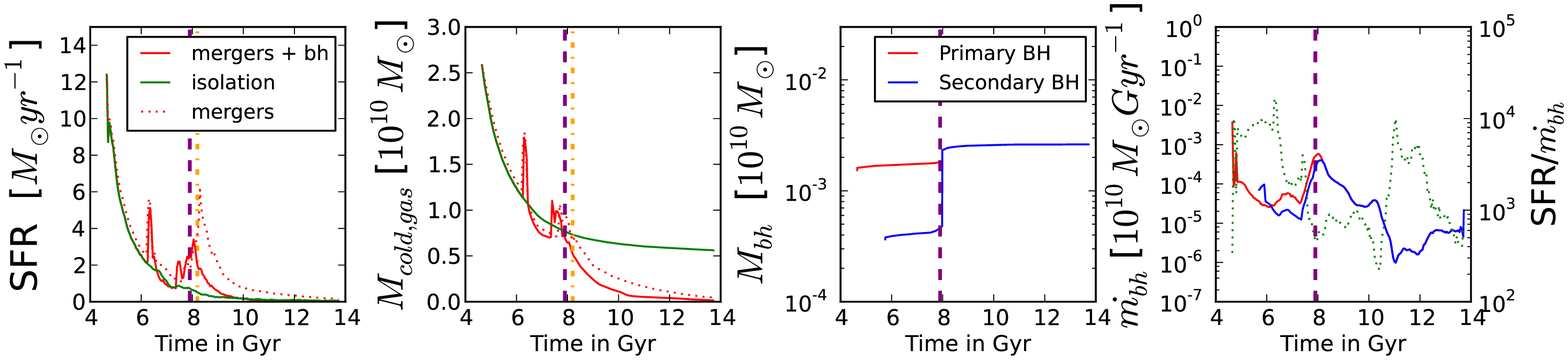,width=.9\textwidth}
\psfig{figure=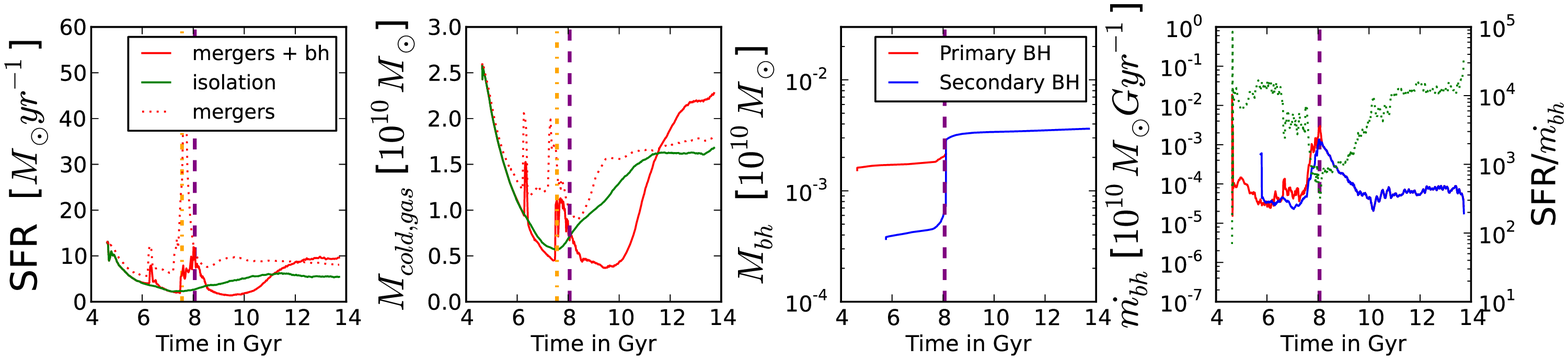,width=.9\textwidth}
\caption{Merger tree 990 with BHs but without (top)
and with a hot gaseous halo (bottom). For each row, the 
panels represent the following: in
  the left plots the SFR of the merger tree is shown,   
while the second plot from the left shows the mass of cold
gas present in the galaxy. The right two panels show the BH mass and
accretion rate. In the left two plots the solid red line shows the primary
galaxy during the merging process, and the green line is the primary
galaxy in isolation. Added to these plots are the results when BH
feedback is not included, represented by the dotted red line. The red
(blue) line in the right two plots shows the properties for the BH in
the center of the primary (secondary) galaxy. When the galaxies and
BHs merge the results for the merged BH continue as a single blue
line.  In the right plots we overplot a green dotted line,
which corresponds to the ratio of SFR and $\dot{m}_{\rm bh}$ (merging 
galaxy), as indicated on the right vertical axis.
The vertical lines in all panels indicate the time when the galaxies
merge (orange dash-dotted for the simulation without BH, and vertical
dashed purple for the simulation with a BH). \label{fig:990bh_cold} }
\end{figure*}

To test this scenario we resimulated all our merger trees with the
addition of BHs and with our implementation of BH feedback (see
Sec. \ref{sec:hydro}).  We also performed a simulation with BH but
without the hot halo component for three merger trees, namely 780,
872, and 990. This is important to separate the effects of BHs and the
hot halo and directly compare our results to previous studies.
The top panels of Fig. \ref{fig:990bh_cold} shows the results for merger tree 990
without the inclusion of the hot halo component. The upper left panel
shows the SFR for three different simulations: host in isolation with
BH, merger simulation with BH and merger simulation without BH. The
SFR for tree 990 without a hot halo and without BH feedback is also
shown for comparison (the same curve as shown in the left panel of
Fig.~\ref{fig:SFR990}). 
It is clear that the SFR rate is reduced in the presence of a BH, with
the second peak lowered by more than a factor two.

The effect of the BH is even more clear in the lower left panel, where
the mass of the cold gas is shown as a function of time. With the
inclusion of BHs and associated feedback, the cold mass decreases and
quickly becomes negligible. The decrease of the SFR and the cold-gas
mass can be easily linked to the increase of the gas-accretion rate
onto the BH (top right panel in Fig. \ref{fig:990bh_cold}). The
accretion rate onto the BH is directly linked to the feedback energy
output, responsible for heating the gas. These results are very
similar to those of previous studies
\citep[e.g.][]{dimatteo2005,springel2005b,Hayward2014}.

Next we included the presence of a hot gas halo in our initial
conditions for merger 990 and reran it with BHs. The results are
presented in the bottom row of Fig. \ref{fig:990bh_cold}.  
The simulation with BHs shows a similar trend as in the previous 
figure, when it is compared to the simulation without BHs but
with hot halo. The total SFR is reduced and
the second peak of SF (at 8 Gyr) drops by more than a factor 4.  After
the merger, SF is quenched to values as low as 2 \Msun yr$^{-1}$,
while without a BH the SFR was around 10 \Msun yr$^{-1}$.

The inclusion of the hot halo has a strong influence on the later
stage of the evolution of SF and cold-gas mass.  After 10 Gyr the hot
halo is able to substantially increase the cold-gas mass in the center
of the halo through cooling, as is cleary shown in the top middle-left
panel of Fig. \ref{fig:990bh_cold}. Thanks to this new cold gas reservoir,
SF can increase again (contrary to the simulation with BH and no hot
gas) and recovers to a SFR of several solar masses per year at the end
of the simulation.

The higher cold-gas fraction in the run with the hot halo does not
only increase the SF burst after eight Gyrs (with respect to a run
without the hot halo) but also increases the accretion rate onto the
BH as shown in the lower right panel in Fig. \ref{fig:990bh_cold}.  This is
true during the merger, but it is even more evident after ($t>9$
Gyrs), when there is a difference of more than one order of magnitude
in the accretion rate of the simulation with and without the hot gas
component.  These higher $\dot m_{bh}$ values result in a larger BH
mass at the end of the run as shown in the bottom middle right panel of
Fig. \ref{fig:990bh_cold}.

\begin{figure*} 
\psfig{figure=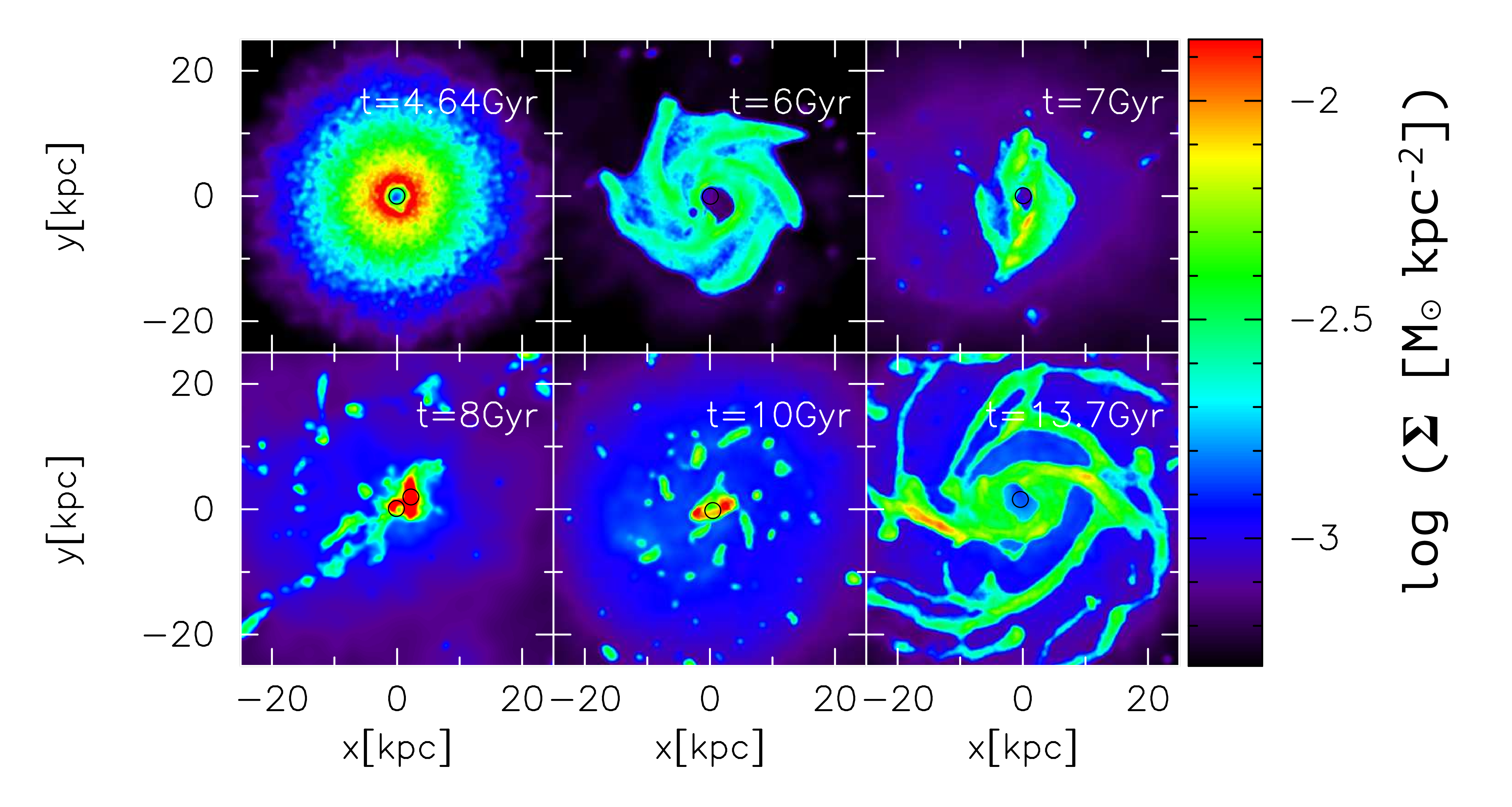,width=\textwidth}
\caption{Gas surface density map of merger tree 990 with BH feedback
  included, shown at six different times. A depletion of gas can be
  seen around the BH (black empty circle) at all times, except during
  the merger at 8 Gyr, when more gas is transported into the nucleus
  of the galaxy, increasing the accretion rate of the
  BH. \label{fig:990bh_map} }
\end{figure*}

The effect of the BH on the gas distribution can be clearly seen in
Fig. \ref{fig:990bh_map} where the surface density map of the gas in
the merger 990 is shown.  Before the merger (t=6 and t=7 Gyr) the BH
is depleting the gas content in the center of the galaxy, while during
the merger (t=8 Gyr) a lot of gas is funnelled into the nucleus.  This
increases the accretion rate by about two orders of magnitude,
releasing a lot of energy from the BH that again heats and depletes
the gas in the central region (t=13.7 Gyr).

The reduced SFR in the simulations including BHs brings the final
galaxies (i.e. the merger products) into better agreement with the
observed stellar-halo mass relation as already shown in
Fig.~\ref{fig:SHM}, in agreement with previous studies that have shown
AGN feedback is necessary to prevent the formation of overly massive
galaxies, even at these mass scales \citep[e.g.][]{somerville2008a}.

Let us now summarise the results of the whole ensemble of simulated
trees with BHs.  The mass of the BHs increases slightly during the
first stage of the simulation, the pre-merger phase. Lower mass BHs
grow relatively faster than larger BHs as a consequence of using the
Bondi Hoyle accretion parametrisation.

The second growth phase is during the merger, when the accretion rate
increases by an order of magnitude in most merger trees (see for
example Fig. \ref{fig:990bh_cold}).  We find a strong correlation between
the presence and size of a starburst and the increase of the accretion
rate of the BHs.  Mergers that produce a large starburst also cause a
large increase in the accretion rate, as both phenomena are related to
the amount of cold gas in the galaxy center.

Most of the BHs merge within the simulation time (this is visible for
example in the middle right panels of Fig. \ref{fig:990bh_cold}, where one
line disappears at around 8 Gyrs). After the merger the accretion rate
is quite low even in the presence of a hot halo component and the mass
of the BH remains roughly constant.

These general trends are in good agreement with previous studies, e.g.
\citet{dimatteo2005}, \citet{springel2005b}, and \citet{Hayward2014},
with the difference that in their simulations SF is quenched
indefinitely because they do not include a hot halo, while in our
simulations we find that after several gigayears the halo has cooled
enough to provide fuel for SF.

\begin{figure} 
\includegraphics[width=0.9\columnwidth]{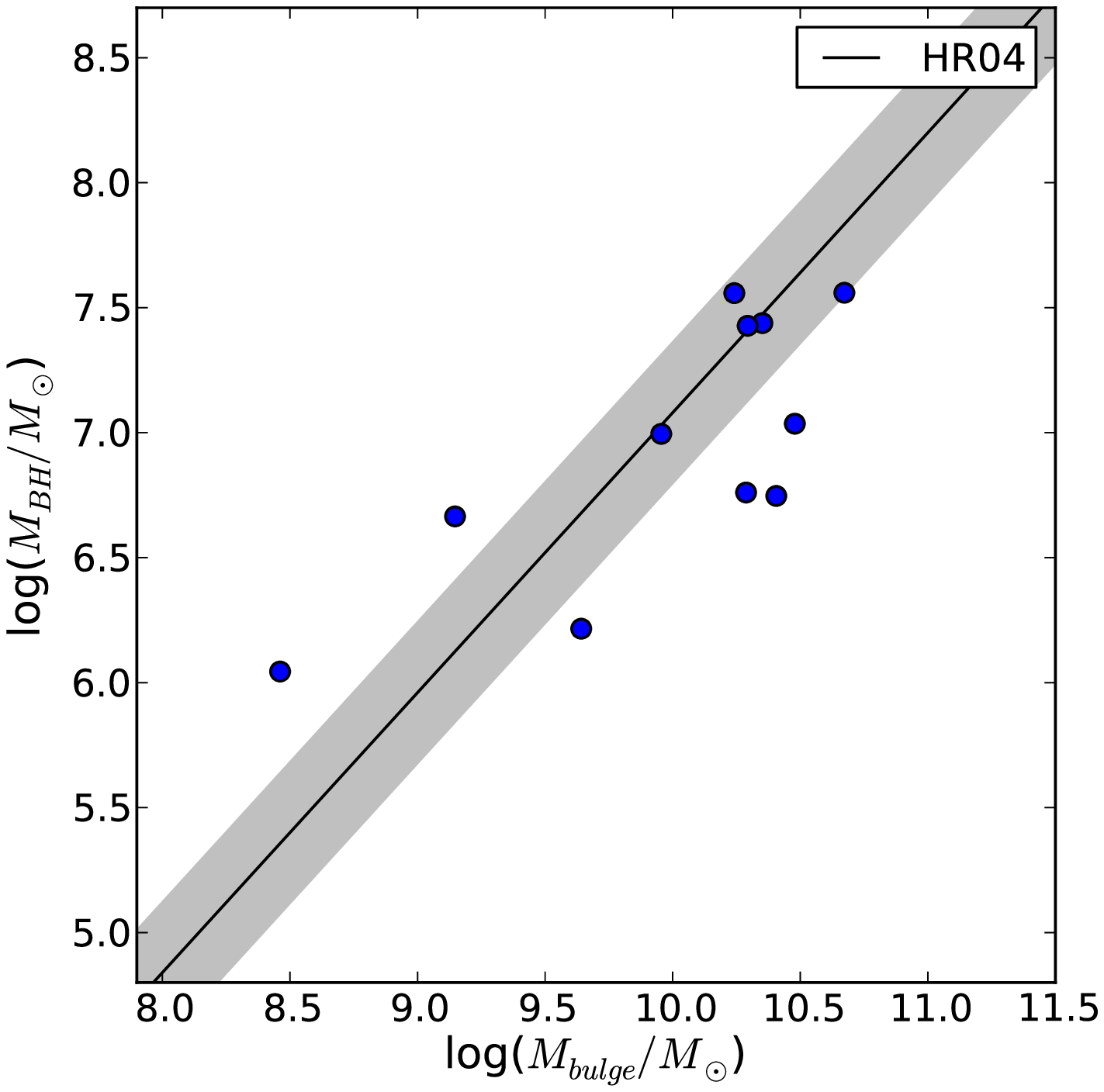}
\caption{Bulge mass versus BH mass for the simulated galaxies at $z=0$
  (blue points).  The black line shows the observed correlation found
  by \citet{haering2004}, while the grey area shows the intrinsic
  scatter in this relation. \label{fig:bulge_bh} }
\end{figure}

The mass of the BHs at the end of the simulation can be compared to
observations, for example using the correlation found between
bulge mass and BH mass
\citep[e.g.][]{haering2004,Sani2011,Kormendy2013}.  In
Fig. \ref{fig:bulge_bh} we show that there is reasonable agreement
between the simulated galaxies and observations. This is due to the
self-regulated nature of the BH growth as shown in previous studies
\citep[e.g.][]{dimatteo2005}.

In two simulations with BHs, namely those of merger trees 1166 and
1684, we see that the above described behaviour does not apply to
every BH.  In tree 1166, the third and smallest BH shows a fast and
constant increase of its mass. In this merger tree there is an
interaction between three galaxies and because of this, the third BH
present in the smallest galaxy is able to move through the 
high-density gaseous medium of the other galaxies. This allows the BH to
constantly accrete gas and keep increasing its mass, until it merges
with the BH in the central galaxy.  In tree 1684, the BH in the
central galaxy shows barely any signatures of a merger and keeps
increasing its mass exponentially throughout the entire
simulation. This is probably due to the large amount of gas accreting
onto the central galaxy.

\begin{table*}
 \begin{tabular}{l|l|l|l|l|l|l|l|l|l}
 \bf Tree nr & $\bf t_{\rm 1}$ & $\bf t_{\rm 2}$ & $\bf
 \mu$& f$_{\rm gas}$ & f$_{\rm bulge, host}$ & f$_{\rm bulge, sat}$ &$\bf M_{\rm burst}$&$\bf M_{\rm cold}$ & $e_{\rm burst}$ \\ 
 & {\it Gyr} & {\it Gyr} & & & & & $10^{10} \Msun$ & $10^{10} \Msun$ &\\ 
\hline
 661  & 5.39  & 6.32  & 6.59 & 0.63 & 0.05 & 0.61 & 0.113 & 1.875 & 0.06\\
 661  & 7.32  & 10.72 & 2.89 & 0.38 & 0.28 & 0.00 & 0.871 & 3.491 & 0.25\\
 872  & 8.96  & 9.35  & 3.59 & 0.20 & 0.07 & 0.62 & 0.017 & 1.820 & 0.01\\
 990  & 7.45  & 8.50  & 2.22 & 0.17 & 0.16 & 0.24 & 0.341 & 1.175 & 0.29\\
 1166 & 7.18  & 8.89  & 3.30 & 0.43 & 0.05 & 0.01 & 0.184 & 1.329 & 0.14 \\
 1166 & 8.00  & 8.89  & 1.62 & 0.27 & 0.18 & 0.03 & 0.101 & 0.854 &  0.12\\ 
 1178 & 7.40  & 9.75  & 2.59 & 0.50 & 0.02 & 0.24 & 0.666 & 2.174 & 0.31\\
 1188 & 6.36  & 8.11  & 1.39 & 0.42 & 0.02 & 0.04 & 0.414 & 1.466 & 0.28\\
 1188 & 10.27 & 10.52 & 8.14 & 0.11 & 0.85 & 0.02 & 0.002 & 0.508 & 0.004\\
 1188 & 12.21 & 13.72 & 2.25 & 0.12 & 0.49 & 0.17 & 0.064 & 0.604 & 0.11\\
 1684 & 8.45  & 8.90  & 2.32 & 0.31 & 0.11 & 0.00 & 0.028 & 2.384 & 0.01\\
 2096 & 7.75  & 10.70 & 1.31 & 0.36 & 0.01 & 0.002 & 0.913 & 2.284 & 0.40\\
 3747 & 5.20  & 6.25  & 4.35 & 0.79 & 0.01 & 0.00 & 0.285 & 1.566 & 0.18\\
 3747 & 9.55  & 11.50 & 3.03 & 0.23 & 0.13 & 0.11 & 0.173 & 2.011 & 0.09
 \end{tabular}
\caption{Simulations with BHs, only nine out of twelve merger trees showed
major mergers. See caption Table
 \ref{table:effwinds}. \label{table:effbh}}
\end{table*}

\begin{figure} 
\includegraphics[width=0.49\textwidth]{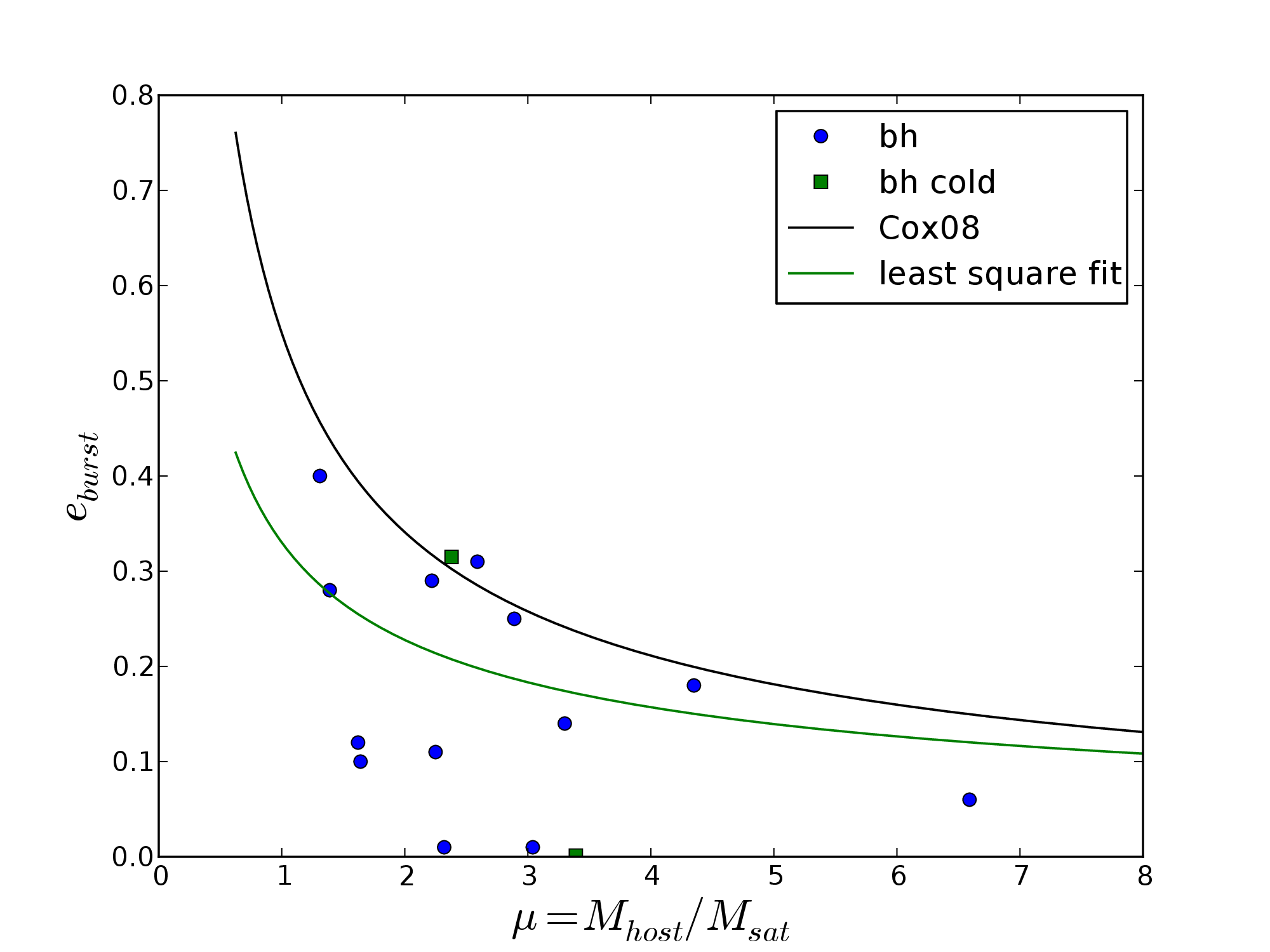}
\caption{Merger mass ratio versus starburst efficiency. The blue points show
 all major merger events in simulations with a BH and a hot
 gaseous halo around the galaxy, while the green squares
show the simulations with black holes and without a hot halo. The black line is the starburst
 efficiency of \citet{cox2008} and the green line represents the
 least squares fit to our results. \label{fig:bursteff_bh} }
\end{figure}

Finally, similar to the previous section, we want to quantify the
ability of mergers in triggering additional SF through the starburst
parameter $e_{\rm burst}$.  The relation between the starburst
efficiency and the merger ratio for simulations with BHs is shown in
Fig. \ref{fig:bursteff_bh}, while Table \ref{table:effbh} is the
analogue of Table \ref{table:effwinds} for these simulations.

It is clear from the figure that the inclusion of both the hot halo
component and BHs strongly reduces the efficiency of massive major
mergers in triggering additional SF with respect to previous studies
\citep[e.g][]{cox2008}.  Both the slope $\gamma$ and the normalisation
$e_{\rm 1:1}$ of the power law are lower than found above, showing
that the correlation between starburst effiency and merger ratio is
much weaker.

The weaker correlation between the merger ratio and the burst
intensity could be partially due to a self-regulation mechanism
between SF and BH feedback.  A massive merger is able to funnel a
significant amount of cold gas towards the nucleus
\citep[e.g.][]{dimatteo2005}; this cold gas provides fuel for the
stellar burst but is also available to be accreted by the BH.  While
in the absence of a central BH a very massive merger will trigger a
strong star formation episode (as shown in figure \ref{fig:bursteff}),
in the presence of BH feedback the merger will also trigger accretion
onto the black hole and the subsequent feedback, which will then
suppress the stellar burst.

\begin{figure} 
\includegraphics[width=0.49\textwidth]{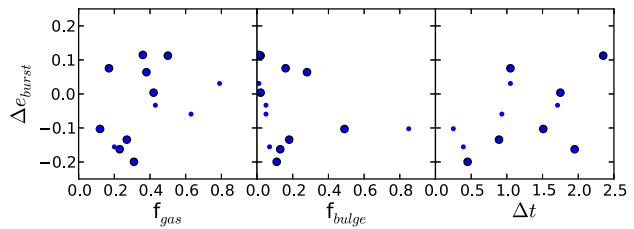}
\caption{Difference between the burst efficiency compared to the power
  law fit to the data in Fig.~\ref{fig:bursteff_bh}, $\Delta e_{\rm
    burst} = e_{\rm burst} - e_{\rm burst, PL}$. The difference is
  shown as a function of the gas fraction $f_{\rm gas}$ (left), the
  bulge fraction of the main galaxy $f_{\rm bulge}$ (middle), and the
  duration of the starburst $\Delta t = t_2 - t_1$ which is a tracer
  of the orbit (right). The big circled points are all major mergers
  with $1<\mu<3$, while the smaller points are all mergers outside of
  this range.\label{fig:bursteff_bh_scat} }
\end{figure}

Finally, the large scatter in the relationship between burst
efficiency and merger mass ratio suggests that other factors can have
an important role in determining the strength of the starburst.  In
Fig. \ref{fig:bursteff_bh_scat} we again investigate if there is a
single parameter that controls this scatter, but we only find a
possible weak trend with $\Delta t$, an indicator of the orbit.  We
plan to address the origin of this scatter in more detail in a future
paper.

\begin{table}
 \begin{tabular}{l|l|l|l|l|l|l|l|l|l}
 \bf Tree & $\bf t_{\rm 1}$ & $\bf t_{\rm 2}$ & $\bf
 \mu$ &$\bf M_{\rm burst}$&$\bf M_{\rm cold}$ & $e_{\rm burst}$ \\ 
 & {\it Gyr} & {\it Gyr} & & $10^{10} \Msun$ & $10^{10} \Msun$ &\\ 
\hline 
 872  & 10.95  & 12.00 & 3.39 & 0.001  & 0.298 & 0.000 \\
 990  & 7.30 & 9.30 & 2.38 &  0.220  & 0.699 & 0.315 
\end{tabular}
\caption{Results of the simulations without a hot halo, but with a black hole. 
Only two
out of twelve merger trees were simulated with these initial conditions.
The columns correspond to columns 1, 2, 3, 4, 8, 9, and 10 from table
 \ref{table:effwinds}. \label{table:cold_effbhs}}
\end{table}

\section{Discussion}
\label{sec:discussion}

This study differs from previous ones in our method of obtaining
initial conditions (IC) for numerical hydrodynamic merger
simulations. Previous studies that included a hot gaseous halo
\citep{cox2006,moster2011a,choi2014} had ad hoc ICs, while here we
adopted ICs that are cosmologically motivated. Therefore, the
progenitors of our galaxies have a broad range of initial properties
(gas fraction, morphology, size, etc.), and consequently, our results
are more representative of an unbiased galaxy population.

\citet{Hayward2014} compared the performance of SPH to {\sc arepo} in
merger simulations. They showed that SPH poorly resolves thermal
instabilities, and it produces a significantly larger hot gaseous halo
around galaxies than {\sc arepo} \citep{Springel2010}, with spurious
blobs of cold gas. The gravitational heating by satellites that we see
in our simulations might be sensitive to the these numerical issues,
and should be more thouroughly tested with grid-based codes or
improved versions of SPH. In addition, the blobs formed in SPH might
spuriously contribute to the heating of the hot halo, although it is
likely a minor effect, as most of the heating occurs shortly before a
merging event.

Although we find clear evidence that the recent merger history should
be taken into account when calculating burst efficiencies, a full
study of the effects of multiple mergers would require an additional
substantial simulation effort, which is beyond the scope of this
paper. We plan to address this in a forthcoming work.
 
In our suite of simulations, we find some significant starbursts for
relatively large merger ratios ($\mu>5$, see Fig. \ref{fig:bursteff}),
contrary to what previous studies found.  Because we focused on major
mergers in this study, our selection criteria limit the number of
minor mergers. Consequently, we do not have a large enough sample of
minor mergers to constrain the dispersion in burst efficiency for
minor mergers.

In Figs. \ref{fig:Temp780} and \ref{fig:multi1}, we see that the
gas mass for the galaxies is nearly constant with time. Because the
stellar mass does increase by a factor of $\sim2$ 
since $z=1$, the gas fractions of these galaxies also decrease by a
factor of 2. Since we are looking at individual galaxies, it is
difficult to compare the gas masses to observations. However,
the trend that the gas fraction decreases by a factor of 2
over this period is compatible with observations \citep[e.g.][]{Santini2014}.
The constant cold gas masses are also in agreement with observations that
the neutral-gas content of galaxies does not vary in galaxies that
 recently underwent a merger \citep{Ellison2015}.

From the stellar mass to halo mass comparison, we find that
overcooling is not a major problem for the merger tree simulations
discussed in this paper. However, this is in part due to the setup of
our simulations, because they are based on SAMs that do include radio
mode feedback. In addition, the rather large value of $\alpha$, the
ratio of specific angular momentum of the hot halo gas and the dark
matter, was calibrated to obtain galaxies with reasonable star
formation histories \citep[see][]{moster2011a}. For a test simulation
with a significantly larger halo mass, we find very high SFRs
caused by overcooling, indicating that at higher masses additional
feedback such as radio mode AGN feedback is required
\citep[e.g.][]{somerville2008a,Fanidakis2011}.

We found, in agreement with previous studies, that the inclusion of
AGN feedback reduces the efficiency of merger-triggered bursts. In our
simulations, black hole accretion and the associated feedback are
modelled in a relatively crude manner. Recent studies
\citep[e.g.][]{choi2014} have shown that the deposition of momentum from
AGN-driven winds originating on the unresolved scales of the nucleus
may have an even more dramatic effect, driving gas out of the nucleus
in high-velocity outflows and further suppressing nuclear star
formation.

\section{Conclusions}
\label{sec:conclusion}

Starbursts triggered by galaxy mergers have long been a topic of great
interest in the literature. Here, we present important progress in the
theoretical understanding of this phenomenon by improving on past
studies in several ways. In particular, we employed the Simulated
Merger Tree approach introduced in MMS14, in which cosmological merger
trees combined with a semi-analytic model are used to obtain the
initial conditions (orbit and galaxy properties) for a sequence of
hydrodynamic merger simulations carried out with the SPH code {\sc
  gadget2}. As motivated by fully cosmological simulations and
observations, but unlike most previous studies, our model galaxy-halo
systems include a hot gaseous halo. We studied the impact of the
inclusion of the hot gas halo, as well as that of BH feedback and
multiple mergers, on triggered star formation enhancements in galaxy
interactions, with a focus on major mergers. 

Our main results can be summarised as follows:

\begin{itemize} 

\item Owing to the presence of the physically and observationally
  motivated hot halo, our simulated galaxies are able to sustain SF
  for longer times, without experiencing artificial strangulation
  due to lack of cold gas.

\item The starburst which results from the merger of two galaxies is
  relatively weak compared with most previous simulations.  This is mostly
  because of the increase of the SFR at later times (due to the hot
  halo), which does not decrease the absolute size of the starburst,
  but does decrease its relative size.

\item In some of our mergers the total SF is {\it decreased} with
  respect to the SF of the same galaxy run in isolation. This implies
  that mergers can also have a {\it negative} effect on SF.  This
  decrease of SF is caused by the transfer of orbital energy of the
  merging galaxies into thermal and kinetic energy of the hot gas and
  by the generation of shocks in the hot halo during the merger.
  These heating processes increase the cooling time of the hot
  circumgalactic gas and lead to a lower SFR in the latest stage of
  the mergers.  This also shows that extra heating terms are present
  during merging processes and should be taken into account in galaxy
  evolution models.

\item We showed that three simultaneously merging galaxies should not
  be treated as two independent and consecutive mergers. The starburst
  resulting from a triple merger can be either similar to the one of
  the merger of the two most massive galaxies or can even be {\it
    smaller}. These results underline once more the complex dynamics
  of multiple mergers and the non-linear effect of different heating
  and cooling mechanisms.

\item When we include thermal feedback from massive central BHs, we
  find that the efficiency of merger-triggered starbursts is reduced,
  especially right after the merger.  Overall, BHs substantially
  reduced the stellar mass of the merger remnants, bringing them in
  better agreement with abundance matching predictions.
  
\item Finally, following previous studies, we described the efficiency
  of mergers in triggering a starburst by a power-law function of the
  merger mass ratio.  In absence of BHs, we find good agreement in the
  normalisation of this relation with \citet{moster2011a} who found
  $e_{1:1} = 0.51$ in their hot halo simulation, while we find
  $e_{1:1}= 0.52$ for a larger sample of galaxies.  Our value for the
  index of the power-law $\gamma = 0.81$ is in reasonable agreement
  with the value found by \citet{cox2008} ($\gamma = 0.69$).  However,
  the scatter that we find is significantly increased, arguing against
  using a single fit to determine burst efficiencies for individual
  galaxies.  When we include feedback from BHs there seems to be a
  weaker dependence of the starburst efficiency on the merger mass
  ratio.

\end{itemize} 

There are still numerous improvements possible for our
simulations. For example, we have a quite crude description of the
thermal feedback from the central black hole and we did not include
momentum deposition from nuclear scale winds \citep{choi2014} or any
jet-like \citep[e.g][]{Cielo2014} or radio feedback from the BH.

Nevertheless our results clearly show the importance of including all
galaxy components (DM, stars, hot gas, cold gas, BH) when performing
galaxy merger simulations, and of adopting cosmologically motivated
merger histories and initial conditions. Our results may be useful for
improving the treatment of merger-driven starbursts in semi-analytic
models of galaxy formation.

\section*{Acknowledgements} 

The authors thank the anonymous referee for his/her useful comments.
The numerical simulations used in this work were performed on the THEO
cluster of the Max-Planck-Institut f\"ur Astronomie at the
Rechenzentrum in Garching.  WK, AVM, and RK acknowledge support from
the Sonderforschungsbereich SFB 881 ``The Milky Way System''
(subproject A1) of the German Research Foundation (DFG).  AVM also
acknowledges financial support to the DAGAL network from the People
Programme (Marie Curie Actions) of the European Union Seventh
Framework Programme FP7/2007-2013/ under REA grant agreement number
PITN-GA-2011-289313. RSS acknowledges the generous support of the
Downsbrough family.


\bibliographystyle{mn2e}
\bibliography{ms}

\appendix
\section{Number of particles in simulations}
\begin{table*}
\begin{tabular}{llllll}
{\bf Tree nr} & {\bf type} & {\bf \# gas particles} & {\bf \# DM particles} & {\bf \# stellar disc particles} & {\bf \# stellar bulge particles}\\
\hline
661 & h  & 1571597 & 334836 & 15085 & 0 \\
661 & s1 & 138912 & 27898 & 2568 & 385 \\
661 & s2 & 742202 & 232175 & 76331 & 0  \\
780 & h & 6280052 & 1349400 &  15564 & 16734  \\
780 & s1 & 5963875 & 1147249 & 120169 &  1574 \\
872 & h & 1453482 & 410260 & 92684 &  6500 \\
872 & s1 & 682182 & 180190 & 29934 & 55610  \\
990 & h & 871822 & 286404 & 59640 &  0 \\
990 & s1 & 560605 & 166434 & 52888 & 14282  \\
1166 & h & 1627114 & 423707 & 43320 &  0 \\
1166 & s1 & 1506889 & 343997 & 42881 & 0  \\
1166 & s2 & 621554 & 147212 & 12450 & 2784  \\
1178 & h & 1537422 & 399421 & 51773 & 0  \\
1178 & s1 & 603668 & 159475 & 30877 & 11323  \\
1188 & h & 588450 &170546  & 43075 & 0  \\
1188 & s1 & 538931 & 129737  &19534  & 3246  \\
1188 & s2 & 202764 & 48912 & 9662 & 0  \\
1188 & s3 & 479564 & 150252 & 36526 &  0 \\
1415 & h & 311220  & 106454 & 6221 & 0  \\
1415 & s1 & 649892 & 142452 & 916 &  0 \\
1415 & s2 & 868248 & 221719 & 38571 &  0 \\
1415 & s3 & 2218312 & 728668 & 299116 &  5864 \\
1684 & h & 1790040 & 394454 & 21398 & 31260  \\
1684 & s1 & 794874 & 231646 & 62530 & 0  \\
2096 & h & 1821984 & 451966 & 58799 & 0  \\
2096 & s1 & 2077989 & 532028 & 99427 & 0  \\
2809 & h & 2415406 & 649315 & 47665 & 0  \\
3747 & h & 1697765 & 412381  & 22361 & 0  \\
3747 & s1 & 405737 & 86786 & 5210 &  0 \\
3747 & s2 & 1344089 & 381888 & 47913 & 10937
\end{tabular}
\caption{The number of particles used to generate galaxies 
in our merger trees. Central galaxies are labelled in the
second column with $h$, while the satellites are 
represented by $s$.
 \label{tab:partnrs}}
\end{table*}

\label{lastpage}

\end{document}